\documentclass[aps,prx,twocolumn,footinbib,floatfix,superscriptaddress]{revtex4-2}
\usepackage{graphicx}
\usepackage{indentfirst}
\usepackage{braket}
\usepackage{float}
\usepackage{amsmath}
\usepackage{physics}
\usepackage{amssymb}
\usepackage{CJK}
\usepackage{esint}
\usepackage{color}
\usepackage[T1]{fontenc}
\usepackage{subfigure}
\usepackage{amsfonts}
\usepackage{footmisc}
\usepackage{scrextend}
\usepackage{multirow}
\usepackage[outdir=./]{epstopdf}

\usepackage[english]{babel}
\usepackage{url}
\usepackage{comment}
\usepackage{array}
\usepackage{subfiles}
\usepackage{tikz}
\usetikzlibrary{shapes,arrows}
\usepackage{mathrsfs}
\usepackage[mathscr]{eucal}

\usepackage[hyperfootnotes=false]{hyperref}
\usepackage{cleveref}
\definecolor{darkblue}{rgb}{0,0,0.5}
\hypersetup{
colorlinks=true,
linkcolor=black,
filecolor=blue,
citecolor=darkblue,  
urlcolor=black,
}

\def\be{\begin{equation}}
\def\ee{\end{equation}}
\def\ba{\begin{eqnarray}}
\def\ea{\end{eqnarray}}
\def\bal{\begin{equation}\begin{aligned}}
\def\eal{\end{aligned}\end{equation}}

\def\bp{\begin{pmatrix}}
\def\ep{\end{pmatrix}}

\usepackage{bm}

\newcommand{\br}{{\bf r}}

\newcommand{\calE}{{\cal E}}
\newcommand{\calF}{{\cal F}}

\newcommand{\1}{^{(1)}}

\usepackage[normalem]{ulem}

\begin{document}

\title{Ultimate accuracy limit of quantum pulse-compression ranging}

\author{Quntao Zhuang}
\email{zhuangquntao@email.arizona.edu}
\affiliation{
Department of Electrical and Computer Engineering, University of Arizona, Tucson, Arizona 85721, USA
}
\affiliation{
James C. Wyant College of Optical Sciences, University of Arizona, Tucson, Arizona 85721, USA
}

\author{Jeffrey H. Shapiro}
\affiliation{Research Laboratory of Electronics, Massachusetts Institute of Technology, Cambridge, Massachusetts 02139, USA}

\date{\today}

\begin{abstract} 
Radars use time-of-flight measurement to infer the range to a distant target from its return's  roundtrip range delay. They typically transmit a high time-bandwidth product waveform and use pulse-compression reception to simultaneously achieve satisfactory range resolution and range accuracy under a peak transmitted-power constraint. Despite the many proposals for quantum radar, none have delineated the ultimate quantum limit on ranging accuracy.  We derive that limit through continuous-time quantum analysis and show that quantum illumination (QI) ranging---a quantum pulse-compression radar that exploits the entanglement between a high time-bandwidth product transmitted signal pulse and and a high time-bandwidth product retained idler pulse---achieves that limit.  We also show that QI ranging offers mean-squared range-delay accuracy that can be 10's of dB better than a classical pulse-compression radar's of the same pulse bandwidth and transmitted energy.
\end{abstract} 

\maketitle

\section{Introduction.}
Classical microwave radars use time-of-flight measurement to infer the range to a distant target from its return's roundtrip range delay $\tau$~\cite{van2001detection1,van2001detection3,mallinckrodt1954optimum,skolnik1960theoretical,skolnik1962introduction,marcum1960statistical,skolnik2008radar}.  Their range-delay resolution $\tau_{\rm res}$, i.e., 
the delay separation needed for reliably distinguishing between two targets based on their range separation, is inversely proportional to the radar pulse's bandwidth $\Delta\omega$. Their ultimate range-delay measurement accuracy for a single target, i.e., the minimum root-mean-squared (rms) estimation error $\delta\tau_{\rm min}$ for localizing a single target, as set by the Cram\'{e}r-Rao bound (CRB), decreases as the signal-to-noise ratio (SNR) increases.  A transform-limited transmitted pulse with duration $T$ and peak power $P_T$, has $\Delta\omega \sim 2\pi/T$ and provides an SNR satisfying ${\rm SNR}\propto P_T T$.  For a radar whose peak power is constrained, these behaviors lead to a conflict between improving the range-delay resolution and improving the ultimate range-delay accuracy.  Using a high time-bandwidth product ($T\Delta\omega\gg 2\pi$) transmitted pulse, however, retains the ${\rm SNR}\propto P_TT$ behavior, but pulse-compression reception results in $\tau_{\rm res} \sim 2\pi/\Delta\omega \ll T$~\cite{skolnik1960theoretical,mallinckrodt1954optimum}.    Range-delay resolution and CRB accuracy, however, are \emph{not} the whole story for radar ranging.  Because range-delay estimation is a nonlinear problem, there  is a minimum SNR below which range-delay performance is significantly worse than the CRB~\cite{zakai1969threshold,chow1981delay,weiss1983fundamental,ianniello1983comparison,weinstein1984fundamental,renaux2008fresh,dardari2009ranging}.  Here, the Ziv-Zakai bound~\cite{zakai1969threshold} affords a useful lower bound on achievable rms accuracy for sub-threshold SNRs.

With the emergence of quantum information science, considerable attention is being paid to the notion of quantum radar~\cite{Lanzagorta,torrome2020introduction,shapiro2020quantum,Sorelli}.  Much of this work has addressed radar operation at optical wavelengths~\cite{giovannetti2001quantum,giovannetti2004,shapiro2007quantum,maccone2020quantum,lloyd2008enhanced}, where background noise has low brightness, i.e., $\ll 1$\,photon/mode, and very low roundtrip radar-to-target-to-radar propagation loss is often assumed.  
Our interest is in microwave radar, where background noise has high brightness, viz., $\sim$\,100's--1000's\,photons/mode, and an unresolved target at range $R$ returns a power that is inversely proportional to $R^4$, making propagation loss severe. Despite this regime's loss and noise, quantum illumination (QI)~\cite{tan2008quantum,Barzanjeh,zhuang2017optimum,shapiro2020quantum} has shown that entanglement offers a 6 dB advantage---over its best classical competitor of the same transmitted energy---in the error-probability exponent for detecting the presence of an unresolved target at a particular location.  

Recently, QI's hypothesis-testing approach was applied to the task of determining which of many contiguous range-delay resolution bins contains a target that is known to be present in one of them~\cite{zhuang2021quantum}.  That first step toward understanding QI's ranging performance did not address QI's ultimate range-delay accuracy, as set by the quantum CRB~\cite{Helstrom_1976,Yuen_1973,Holevo_1982} at high SNR and the quantum Ziv-Zakai bound (ZZB)~\cite{tsang2012ziv} in the sub-threshold SNR region.  This Letter will remedy those deficiencies by developing a continuous-time framework for QI's entanglement-assisted range-delay estimation and comparing its predictions to corresponding results for classical, i.e., coherent-state, radar.

Our proposed QI ranging is a quantum pulse-compression radar that benefits from the entanglement between a high time-bandwidth product transmitted signal pulse and a high time-bandwidth product retained idler pulse.  In comparison to a classical pulse-compression radar of the same bandwidth and transmitted energy, our quantum CRB analyses show that QI's mean-squared accuracy above its SNR threshold is 3\,dB better than the corresponding above-threshold classical performance.  QI's 6\,dB advantage in error-probability exponent over classical radar in determining the target's range-resolution bin~\cite{zhuang2021quantum}, however, provides a 6\,dB reduction in its SNR threshold relative to that of classical radar.  Remarkably, this threshold reduction translates into an entanglement-assisted mean-squared accuracy that can be 10's of dB better than classical performance at the same SNR.

\section{Quantum description of range-delay estimation}
The quantum range-delay estimation problem is as follows.  The radar transmits a single spatial-mode field characterized by a photon-units, positive-frequency field operator 
\be
\hat{E}_S(t) =  \int\!\frac{{\rm d}\omega}{2\pi}\,\hat{A}_S(\omega)e^{-i(\omega_0+\omega)t},
\label{ES}
\ee
where $\omega_0$ is the carrier frequency~\cite{footnote1}.  In both our classical and quantum pulse-compression radars, this field operator's excitation will have duration $T$, bandwidth $\Delta\omega$ satisfying $2\pi/T \ll \Delta\omega \ll \omega_0$, and average photon number $
\calE =  \int\!{\rm d}t\,\langle\hat{E}_S^\dagger(t)\hat{E}_S(t)\rangle.$  
From an unresolved, nonfluctuating target at range $R$, the radar receives a photon-units, positive-frequency field operator $\hat{E}_R(t)$ given by
\begin{align} 
&\hat{E}_R(t)=\sqrt{\kappa}\,e^{i\theta_R}\hat{E}_S(t-\tau)+\sqrt{1-\kappa}\,\hat{E}_B(t), 
\label{ER}
\end{align}
where: $\kappa$ is the roundtrip radar-to-target-to-radar transmissivity; $\theta_R$ is the phase shift incurred in reflection from the target; $\tau = 2R/c$, with $c$ being light speed, is the target's range delay; and 
\be
\hat{E}_B(t) = \int\!\frac{{\rm d}\omega}{2\pi}\,\hat{A}_B(\omega)e^{-i(\omega+\omega_0)t}
\ee
is the background radiation's field operator.  In keeping with a microwave radar's interrogating a distant unresolved target, we shall assume that the target is known to lie in the range uncertainty interval $\mathcal{R} = [R_{\rm min}, R_{\rm max}]$ with $\Delta R \equiv R_{\rm max}-R_{\rm min}\ll (R_{\rm min} + R_{\rm max})/2$. It then follows that the range-delay $\tau$ will lie in $[\tau_{\rm min}, \tau_{\rm max}]$ with $\tau_{\rm min} = 2R_{\rm min}/c$,  $\tau_{\rm max} = 2R_{\rm max}/c$, and $\Delta \tau \equiv \tau_{\rm max}-\tau_{\rm min} = 2\Delta R/c $.  Also, $\kappa$ will be approximately constant over the range uncertainty interval and satisfy $\kappa\ll 1$.  The background radiation---at least over $\hat{E}_S(t)$'s excitation bandwidth---is in a thermal state whose average photon number per mode is $N_ B/(1-\kappa) \approx N_B$, where
\be 
N_B=1/[\exp\!\left(\hbar \omega_0 /k_B T_B\right)-1] \gg 1,
\label{T_radar}
\ee 
with $\hbar$ being the reduced Planck constant, $k_B$ the Boltzmann constant, and $T_B$  the radar receiver's noise temperature.
The range-delay estimation task is to make a minimum rms error estimate of $\tau$ from a measurement made on $\{\hat{E}_R(t) : t\in \mathcal{T}\}$, where $\mathcal{T}$ includes all times for which there could be any target return from the range uncertainty interval.

Before proceeding further, there is an important point to make about $\theta_R$ and $\omega_0$ that is revealed by the frequency domain version of 
Eq.~\eqref{ER}, viz., 
\be 
\hat{A}_R(\omega)=\sqrt{\kappa}\,e^{i[(\omega_0 +\omega) \tau+\theta_R]} \hat{A}_S(\omega)+\sqrt{1-\kappa}\,\hat{A}_B(\omega).
\label{phase_coherent}
\ee 
If $\theta_R$ is modeled as uniformly distributed on $[0,2\pi]$, as is typically the case, the $\omega_0\tau$ term in Eq.~\eqref{phase_coherent} becomes uninformative and the radar must implement a measurement that is not destroyed by phase randomness.  (In classical radar this task is accomplished by means of envelope detection after matched filtering.) On the other hand,
if $\theta_R$ is known, the ensuing range-delay ambiguities spaced $2\pi/\omega_0$ apart~\cite{skolnik1960theoretical} prevent the $\omega_0\tau$ term in Eq.~\eqref{phase_coherent} from being useful. That said, we shall set $\theta_R = 0$ and $\omega_0 = 0$ in Eq.~\eqref{phase_coherent} in evaluating the quantum CRB and ZZB, recognizing, by convexity, that the results obtained therefrom are lower bounds on their phase-incoherent (random $\theta_R$) counterparts. See Appendix~\ref{app:phase_noise} for evidence supporting the minimal impact, on classical radar, of assuming $\theta_R =0$ when $\omega_0=0$. 

Our rms accuracy ($\delta \tau$) assessments for the classical and QI pulse-compression radars combine the quantum CRB~\cite{Helstrom_1976,Yuen_1973,Holevo_1982} and the quantum ZZB~~\cite{tsang2012ziv}.  Specifically,  when the radar in question has its SNR
above its range-delay estimation's SNR threshold, we use $\delta \tau \approx \delta \tau_{\rm CRB} = 1/\calF_\tau$, where $\calF_\tau$ is the Fisher information about $\tau$ contained in $\{\hat{E}_R(t) : t \in \mathcal{T}\}$ (see Appendix~\ref{app:fisher}),
\be
\mathcal{F}_\tau = \lim_{{\rm d}\tau\rightarrow 0}
\frac{8\!\left(1-\sqrt{{\rm tr}\!\left[\!\left(\sqrt{\hat{\rho}_\tau}\,\hat{\rho}_{\tau+{\rm d}\tau}\sqrt{\hat{\rho}_\tau}\right)^2\right]}\right)}{{\rm d}\tau^2},
\ee
where $\hat{\rho}_u$ is the state of the classical radar's received field---and, for the quantum radar, the joint state of its received and retained fields---when the range delay is $u$.  
Alternatively, when the radar's SNR is below threshold, we use $\delta \tau \approx \delta\tau_{\rm ZZB}$, where
\begin{align}
\delta \tau_{\rm ZZB}=
\sqrt{\int_0^{\Delta\tau}\!{\rm d}\tau'\, \tau'\!\left(1-\frac{\tau'}{\Delta\tau}\right)\!P_e(\tau')},
\label{Ziv_Zakai_simplified}
\end{align}
with $P_e(\tau')$ being the minimum error probability---from the likelihood-ratio test for the classical radar~\cite{van2001detection1} and from the Helstrom limit~\cite{Helstrom_1976} for the quantum radar---for distinguishing between the equally-likely hypotheses $H_0$ = target present at range delay $\tau_{\rm min}$ and $H_1$ = target present at range delay $\tau_{\rm min}+\tau'$.  As will be discussed below, however, we will use the quantum radar's Chernoff bound~\cite{zhuang2021quantum,Audenaert2007,Pirandola2008} in lieu of the Helstrom limit, because the former is easily obtained whereas the latter is not (see Appendix~\ref{app:ziv-zakai}).  We expect this substitution will have a modest effect on our results, because the quantum Chernoff bound is known to be exponentially tight in $P_e(\tau')$'s SNR dependence.

\section{Classical pulse-compression radar}
Our classical pulse-compression radar will emit $\hat{E}_S(t)$ in the coherent state $|\sqrt{\calE}\,{\bf s}(t)e^{-i\omega_0t}\rangle$, where 
\be {\bf s}(t) = (2\pi T^2)^{-1/4}\exp\!\left(-t^2/4T^2 +i\Delta\omega t^2/2T\right),
\label{ChirpedGaussian}
\ee
with $2\pi/T \ll \Delta\omega \ll \omega_0$.  Physically, this is a narrowband, but high time-bandwidth product, chirped-Gaussian pulse with average photon number $\calE$ and rms time duration $T$.  Moreover, because
\be
S(\omega) \equiv \int\!{\rm d}t\,{\bf s}(t)e^{-i\omega t} \approx \frac{\exp\!\left(-\omega^2/4\Delta\omega^2\right)}{(\Delta\omega^2/2\pi)^{1/4}},
\ee
so that $\int\!\frac{{\rm d}\omega}{2\pi}\,\omega^2|S(\omega)|^2 = \Delta\omega^2,$
we thus have that $\Delta\omega$ is ${\bf s}(t)$'s rms bandwidth.  The quantum CRB---with $\theta_R = 0$ and $\omega_0 = 0$---for this classical radar's rms range accuracy with quantum-optimal reception is (see Appendix~\ref{app:fisher})
\begin{align} 
&\delta \tau_{\rm CRB}^{\rm C}=\frac{1}{\Delta\omega\sqrt{2\kappa \calE/(N_B+1/2)}} \approx  \frac{1}{\Delta\omega \sqrt{2\,{\rm SNR}}},
\label{dt_C}
\end{align} 
where ${\rm SNR} \equiv \kappa \calE/N_B$ and $N_B \gg 1$ as is typical for microwave operation.

Interestingly, a semiclassical treatment of a radar using this chirped-Gaussian pulse transmitter and ideal heterodyne reception results in a \emph{classical} CRB that gives 
the same result as Eq.~\eqref{dt_C} for $N_B \gg 1$.  Moreover, for $N_B \gg 1$, both the quantum CRB and the  classical CRB for heterodyne reception match well-known classical results~\cite{slepian1954estimation,mallinckrodt1954optimum,skolnik1960theoretical, kotel1960theory,woodward2014probability} for the high-SNR range-delay accuracy.  Furthermore, convexity implies that no classical-state transmitter can outperform the best coherent-state radar, and the CRB from Eq.~\eqref{dt_C} holds for all pulse shapes with rms bandwidth $\Delta\omega$ (see Appendix~\ref{app:fisher}).  Thus we conclude that a coherent-state transmitter with ideal heterodyne reception is the quantum optimum classical-state radar for above-threshold microwave range-delay estimation when $\kappa \ll 1$ and $N_B\gg 1$.  

In contrast to the CRB---which bounds the range-delay accuracy for estimating an unknown, non-random $\tau$---the ZZB is a Bayesian result that assumes $\tau$ to be uniformly distributed on $\tau \in [\tau_{\rm min}, \tau_{\rm max}]$. Consequently, as ${\rm SNR} \rightarrow 0$, we have $\delta\tau_{\rm ZZB}$ will approach the range-delay distribution's standard deviation, $\sigma_\tau = \sqrt{\Delta \tau^2/12}$, which is typically much greater than the radar's range-delay resolution $\tau_{\rm res}$.  Using well-known results (see Appendix~\ref{app:ziv-zakai})---and $\kappa \ll1, N_B\gg 1$---we have that the classical radar's likelihood-ratio test for ideal heterodyne detection results in
\begin{align}
P_e(\tau') &= Q\!\left[\sqrt{{\rm SNR}(1-e^{-\Delta\omega^2\tau'^2/2})}\right]\\[.05in]
&\le \exp[-{\rm SNR}(1-e^{-\Delta\omega^2\tau'^2/2})/2]/2,
\end{align}
where $Q(x) \equiv \int_x^\infty\!{\rm d}y\,e^{-y^2/2}/\sqrt{2\pi}$, and the upper bound is the both the classical Chernoff bound for ideal heterodyne reception and the quantum Chernoff bound for the quantum-optimum measurement made on $\{\hat{E}_R(t) : t\in \mathcal{T}\}$.  The Chernoff bound gives 
$\delta\tau_{\rm ZZB\text{-}QCB}^{\rm C} \sim \sigma_{\tau}e^{-{\rm SNR}/4}$ 
at low SNRs, but at high SNRs we find~(see Appendix~\ref{app:ziv-zakai}) $
\delta\tau_{\rm ZZB}^{\rm C} \rightarrow \delta\tau_{\rm CRB}^{\rm C}$ 
for the exact error probability, whereas $
\delta\tau_{\rm ZZB\text{-}QCB}^{\rm C}\rightarrow \sqrt{2}\,\delta\tau_{\rm CRB}^{\rm C}.$  
The latter behavior is not surprising:  at high SNRs Eq.~\eqref{Ziv_Zakai_simplified} is dominated by contributions from small values of $\tau$, for which the Chernoff bound is not sufficiently accurate in approximating $P_e(\tau')$ to recover the CRB~(see Appendix~\ref{app:ziv-zakai}).

Although we have chosen the chirped-Gaussian pulse of Eq.~\eqref{ChirpedGaussian} for analytical convenience, the quantum CRB from Eq.~\eqref{dt_C} and the ZZB's high-SNR and low-SNR asymptotes apply to all pulse shapes with rms bandwidth $\Delta\omega$~(see Appendices~\ref{app:fisher} and \ref{app:ziv-zakai}).

\section{Quantum pulse-compression radar}
Our quantum pulse-compression radar will use a continuous-wave, frequency-degenerate, spontaneous parametric downconverter to produce signal and idler beams whose photon-units, positive-frequency field operators, $\hat{E}_S(t)$ from Eq.~\eqref{ES} and 
\be
\hat{E}_I(t) =  \int\!\frac{{\rm d}\omega}{2\pi}\,\hat{A}_I(\omega)e^{-i(\omega_0-\omega)t},
\ee 
are in a zero-mean jointly-Gaussian state characterized by their nonzero Fourier-domain correlations:  
\be
\langle \hat{A}_K^\dagger(\omega)\hat{A}_K(\omega')\rangle  = 2\pi S^{(n)}(\omega)\delta(\omega-\omega'),
\ee 
for $K = S,I$,
and 
\be
\langle \hat{A}_S(\omega)\hat{A}_I(\omega')\rangle = 2\pi S^{(p)}(\omega)\,\delta(\omega-\omega').
\ee   
The quantum radar transmits a $T$-s-long pulse of its signal beam, where $T \gg 2\pi/\Delta\omega$, while retaining the companion idler pulse for a joint measurement with $\{\hat{E}_R(t): t \in \mathcal{T}\}$.

For analytical convenience, we will take signal and idler's phase-insensitive (fluorescence) spectrum to be
\be
S^{(n)}(\omega)/2\pi = \frac{N_Se^{-\omega^2/2\Delta\omega^2}}{\sqrt{2\pi}},
\label{SnOmega}
\ee
making their average photon number, $
\int_0^T\!{\rm d}t\,\langle\hat{E}^\dagger_S(t)\hat{E}_S(t)\rangle= T\!\int\!\frac{{\rm d}\omega}{2\pi}\,S^{(n)}(\omega) =N_S \Delta \omega T= \mathcal{E},$ and mean-squared bandwidth  $T\!\int\!\frac{{\rm d}\omega}{2\pi}\,\omega^2 S^{(n)}(\omega)/\calE=\Delta \omega^2$ match that of our classical pulse-compression radar, and we will take the signal and idler's phase-sensitive cross spectrum to be
\be
S^{(p)}(\omega) = \sqrt{S^{(n)}(\omega)(S^{(n)}(\omega) + 1)},
\ee
making their quadratures maximally entangled~\cite{shapiro2020quantum}. Note that the assumption of Gaussian fluorescence spectra is not essential for the quantum advantages derived below to hold~(see Appendices~\ref{app:fisher} and \ref{app:ziv-zakai}).


As is well known for QI target detection~\cite{shapiro2020quantum}, our quantum pulse-compression radar's performance advantage will come from its phase-sensitive cross spectrum, $S^{(p)}(\omega) $, greatly exceeding the classical limit on that cross spectrum, $S^{(n)}(\omega)$, when  $N_S/\sqrt{2\pi} = \max_\omega \!\left[S^{(n)}(\omega)\right] \ll 1$.  This low-brightness condition implies that our quantum radar will need a much longer pulse duration than a high-brightness classical competitor of the same bandwidth and energy.  

The quantum pulse-compression radar's Fisher information is easily evaluated~(see Appendix~\ref{app:fisher}), leading---when $N_S \ll 1$, $\kappa \ll 1$, and $N_B\gg1$---to the range-delay CRB  
\be
\delta \tau^{\rm Q}_{\rm CRB} = 1/2\Delta\omega\sqrt{\rm SNR} = \delta\tau^{\rm C}_{\rm CRB}/\sqrt{2},
\label{dt_E}
\ee 
a result that can be shown to be the best high-SNR performance of \emph{all} entanglement-assisted radars~(see Appendix~\ref{app:fisher}).   When $N_S \gg 1$, however, the preceding quantum advantage vanishes, as expected from the previous paragraph, and we get~(see Appendix~\ref{app:fisher})
$\delta \tau^{\rm Q}_{\rm CRB}  = \delta\tau^{\rm C}_{\rm CRB}.$

To find the ZZB for our quantum radar we will focus on the $N_S \ll 1, \kappa\ll 1, N_B\gg 1$ regime and use the quantum Chernoff bound (QCB) for range-bin discrimination that applies when $T \gg \Delta \tau$~\cite{zhuang2021quantum}(see Appendix~\ref{app:ziv-zakai}), as will be necessary for our quantum radar to reach its threshold SNR, viz.,
\be
P_e(\tau')\le \exp[-2\,{\rm SNR}(1-e^{-\Delta\omega^2\tau'^2/2})]/2,
\ee
in place of the (challenging to obtain) exact error probability.  At high SNRs we get
$\delta \tau_{\rm ZZB\text{-}QCB}^{\rm Q} \approx \delta \tau_{\rm CRB}^{\rm Q},$
whereas 
at low SNRs we find~(see Appendix~\ref{app:ziv-zakai})
$\delta\tau _{\rm ZZB\text{-}QCB}^{\rm Q} \approx\sigma_\tau e^{-{\rm SNR}}.$ 
Comparing to the below-threshold classical result, $\delta\tau_{\rm ZZB\text{-}QCB}^{\rm C} \approx \sigma_\tau e^{-{\rm SNR}/4}$, shows the beneficial effect of QI's 6\,dB higher error-probability exponent~\cite{tan2008quantum,zhuang2021quantum}.  

Strictly speaking, using an error-probability \emph{upper} bound in an expression that provides a range-delay accuracy \emph{lower} bound is no longer guaranteed to provide a lower bound.  However, because $\delta\tau_{\rm ZZB\text{-}QCB}^{\rm Q}$ approaches the quantum CRB at high SNR and converges to $\sigma_\tau$ as ${\rm SNR} \rightarrow 0$, we believe it to be a good approximation to the  $\delta \tau$ of our quantum radar~(see Appendix~\ref{app:ziv-zakai}).  

\begin{figure}[t]
    \centering
    \includegraphics[width=0.475\textwidth]{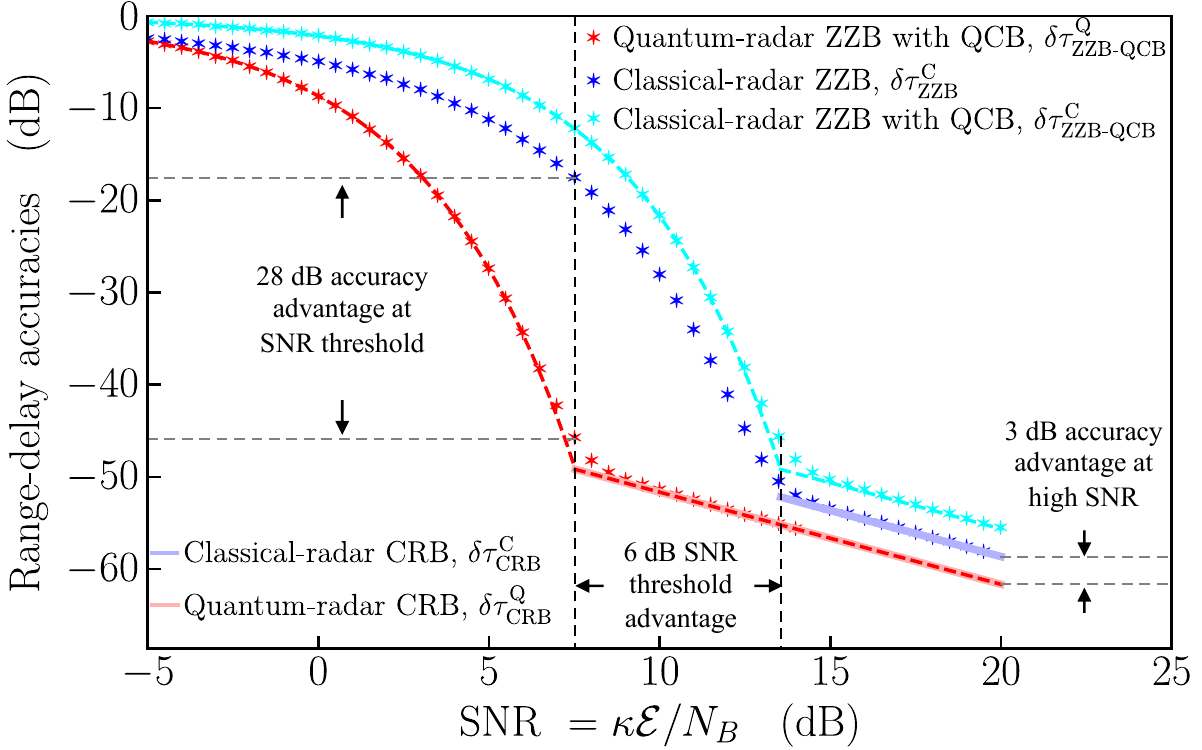}
    \caption{Normalized mean-squared range-delay accuracies in dB,  $20\log_{10}(\delta\tau/\sigma_\tau)$, versus SNR in dB.  The plots assume $\Delta\omega/2\pi = 10^6\,$Hz and $\Delta R = 5$\,km.  Results are shown, top to bottom, for $\delta\tau_{\rm ZZB\text{-}QCB}^{\rm C}$ (cyan stars), $\delta\tau_{\rm ZZB}^{\rm C}$ (blue stars), and $\delta\tau_{\rm ZZB\text{-}QCB}^{\rm C}$ (red stars).  Also plotted, in the corresponding colors, are their high-SNR and low-SNR asymptotic behaviors, along with vertical dashed lines showing their SNR thresholds, ${\rm SNR}_{\rm thresh}^{\rm C}$ and ${\rm SNR}_{\rm thresh}^{\rm Q}$, computed from the intersections of each radar's high-SNR and low-SNR asymptotic behaviors.
    \label{fig:dR}
    }
\end{figure}

\section{Accuracy comparison}
Figure~\ref{fig:dR} presents an example that illustrates the range-delay advantages provided by QI's entanglement-assisted pulse-compression radar over classical pulse-compression radar.  There, we plot results for normalized mean-squared range-delay accuracies versus the radar's SNR, both in dB.  Here we see that the ZZBs (stars) all show a clear threshold phenomenon, as predicted from their high-SNR and low-SNR asymptotic results (dashed lines):  $\delta\tau_{\rm CRB}^{\rm Q} = 1/2\Delta\omega\sqrt{\rm SNR}$ and $\delta\tau_{\rm CRB}^{\rm C}=1/\Delta\omega\sqrt{2\,{\rm SNR}}$ at high SNR, and   
$\delta\tau_{\rm ZZB\text{-}QCB}^{\rm Q} = \sigma_\tau e^{-{\rm SNR}}$ and  $\delta\tau_{\rm ZZB\text{-}QCB}^{\rm C} = \sigma_\tau e^{-{\rm SNR}/4}$ at low SNR.  
For each radar, the threshold signal-to-noise ratio, ${\rm SNR}_{\rm thresh}$, at which range-delay accuracy diverges from the CRB can be obtained by matching its low-SNR and high-SNR $\delta\tau_{\rm ZZB\text{-}QCB}$ asymptotes. We find that~(see Appendix~\ref{app:ziv-zakai})
\be
{\rm SNR}_{\rm thresh}^{\rm Q}={\rm SNR}_{\rm thresh}^{\rm C}/4 = f(1/2\Delta\omega^2\sigma_\tau^2)/2,
\label{SNR_threshold}
\ee 
where $x=f(y)$ is the inverse function of $y=x e^{-x}$. These thresholds are shown by vertical dashed lines in Fig.~\ref{fig:dR}. They match well to the numerical results (red stars for the quantum radar and cyan or blue stars for the classical radar) found by evaluating Eq.~\eqref{Ziv_Zakai_simplified}.  As predicted by Eq.~\eqref{SNR_threshold}, the quantum radar's SNR threshold is 6\,dB lower than that of the classical radar, as highlighted in Fig.~\ref{fig:dR}. This 6\,dB advantage from using entanglement has been verified numerically for a variety of $\Delta \tau$ values~(see Appendix~\ref{app:ziv-zakai}).

Remarkably, for a $\Delta\omega/2\pi=10^6$\,Hz rms-bandwidth quantum radar interrogating a target located within a $\Delta R=5$\,km range uncertainty, operation at ${\rm SNR} = {\rm SNR}_{\rm thresh}^{\rm Q}$ provides a 28\,dB advantage in mean-squared range-delay accuracy compared to a classical radar of the same bandwidth and pulse energy.  More generally, asymptotic analyses~(see Appendix~\ref{app:ziv-zakai}) show that the quantum radar's mean-squared accuracy advantage when operating at ${\rm SNR}_{\rm thresh}^{\rm Q}$ obeys 
\be
(\delta \tau_{\rm ZZB\text{-}QCB}^{\rm C})^2/(\delta \tau_{\rm ZZB\text{-}QCB}^{\rm Q})^2
\sim (\Delta\omega \sigma_\tau)^{3/2},
\label{adv_asym}
\ee
and thus grows with both increasing bandwidth and range-delay uncertainty. 

\section{Discussion}
Microwave radar is a challenging venue for exploiting quantum entanglement~\cite{jonsson2021quantum}. In this Letter, we developed a continuous-time treatment of QI target ranging and compared its performance to that of a classical  pulse-compression radar.  Both use time-of-flight measurements to infer target range,   hence both have SNR thresholds below which their range-delay measurement accuracy is far worse than their CRB limit.  Our quantum radar has a 6\,dB lower threshold SNR than the classical radar.  Consequently, when the quantum radar operates at its threshold SNR, its mean-squared range-delay accuracy can be 10's of dB better than its classical competitor's. Although we have
yet to identify a receiver design for achieving our quantum
radar's range-delay accuracy advantage, such a large
advantage is a much better prospect for retaining a significant
advantage with a practical, but sub-optimal, receiver than is the case for QI target detection.

\begin{figure}[t]
    \centering
    \includegraphics[width=0.35\textwidth]{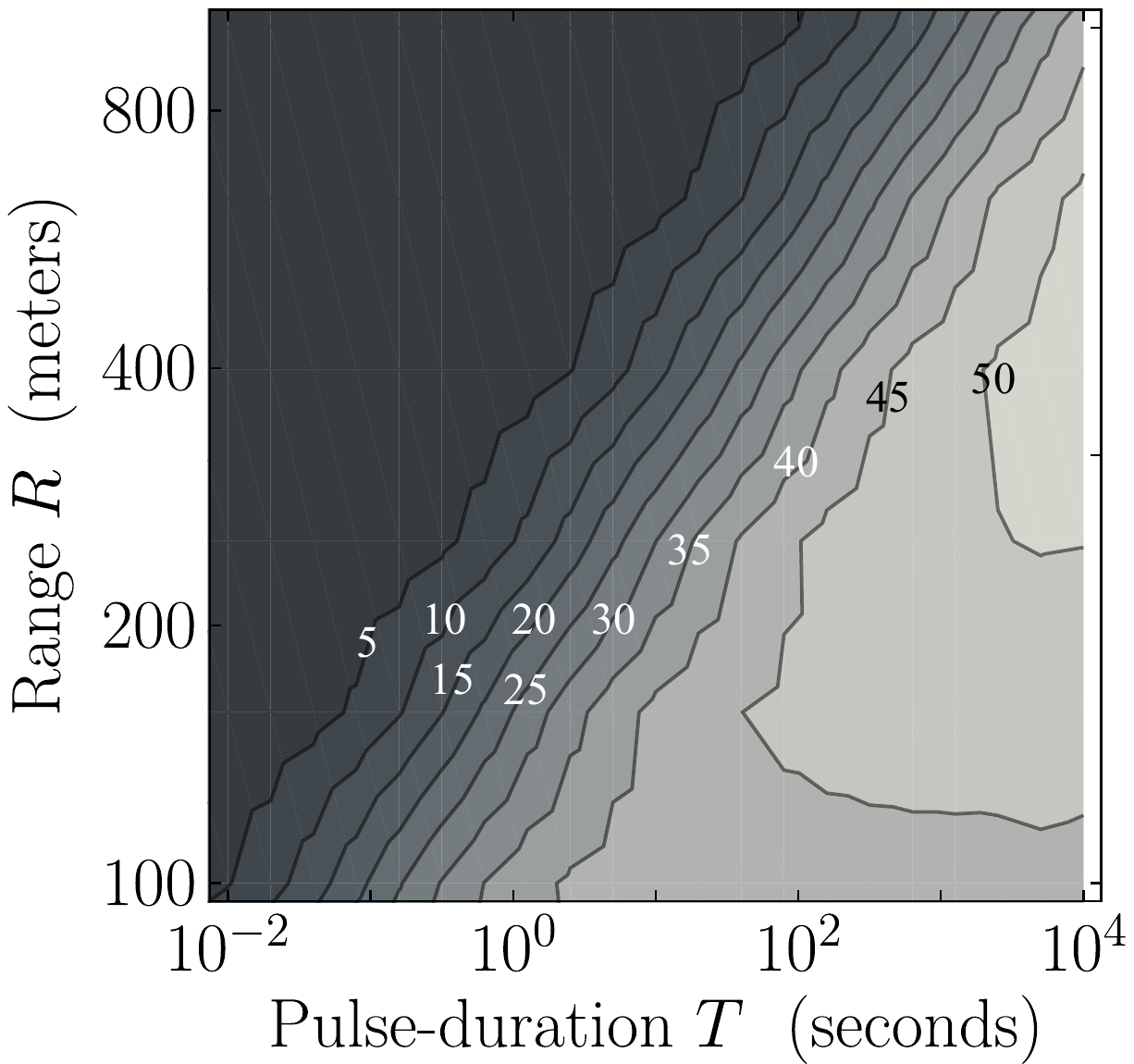}
    \caption{
    Contour plot of mean-squared range-delay accuracy advantage (in dB) at ${\rm SNR}_{\rm thresh}^{\rm Q}$ assuming range uncertainty $\Delta R=R/100$ at range $R$.
    \label{fig:radar_scenario}
    }
\end{figure}

Putting aside the task of developing a practical receiver for QI target ranging, any such system will be subject to the same difficulties~\cite{shapiro2020quantum,jonsson2021quantum,shapiro2021microwave} that preclude QI's utility for long-range target detection unless exceedingly long pulse durations can be employed so that ${\rm SNR} = \kappa MN_S/N_B$ values---where $M$ is the transmitter's time-bandwidth product~\cite{footnote2}---on the order of 5-to-10\,dB can be achieved for $N_S \ll 1$, $\kappa \ll 1$, and $N_B \gg 1$. 

To illustrate the trade-off between range-delay quantum advantage and pulse duration, consider a $W$-band ($\omega_0/2\pi = 100\,$GHz) radar used to localize a tiny unmanned aerial system (UAS).  
$W$-band radar is both high precision and robust to degraded visual environments~\cite{Wband}, hence it is widely applicable to UAS detection~\cite{UAS}.  Figure~\ref{fig:radar_scenario} is a contour plot of this radar's quantum advantage in mean-squared range-delay accuracy for a $\Delta R = R/100$ range uncertainty when operated at ${\rm SNR}_{\rm thresh}^{\rm Q}$.  It assumes an ideal radar implementation for which $\kappa = (G_T/4\pi R^2)(\sigma A_R/4\pi R^2)$, with  $G_T = A_R/(2\pi c/\omega_0)^2$ being the radar's antenna gain, $A_R = 1\,{\rm m}^2$ its antenna area, and  $\sigma = 0.01\,{\rm m}^2$ the UAS's radar cross-section.  It also assumes a $T_B = 150\,$K noise temperature, giving $N_B = 32$.  These parameters determine the $N_S$ needed to reach ${\rm SNR}_{\rm thresh}^{\rm Q}$, and hence the resulting quantum advantage.  As seen in Fig.~\ref{fig:radar_scenario}, a 0.01\,s pulse duration provides a 20\,dB quantum advantage at $R = 100\,$m, but an impractically long 100\,s pulse duration is needed to realize this same advantage at $R = 1\,$km, because $T/R^4$ must be constant to maintain constant quantum advantage.

In conclusion, recall the meta-lesson of Tan et al.'s QI target detection: an entanglement-based advantage can, in principle, be realized in an entanglement-breaking scenario,  so do not dismiss using entanglement in such a scenario.  Thus, bear in mind our Letter's meta-lesson: an enormous entanglement-based advantage can, in principle, be realized in an entanglement-breaking scenario, so be even less willing to dismiss exploiting entanglement in such a scenario.

Q.Z.\@ acknowledges the Defense Advanced Research Projects Agency (DARPA) under Young Faculty Award (YFA) Grant No.\@ N660012014029, Office
of Naval Research under Grant No.\@ N00014-19-1-2189, Craig M.\@ Berge Dean's Faculty Fellowship of University of Arizona and support from Raytheon Technologies. J.H.S.\@ acknowledges support from the MITRE Corporation's Quantum Moonshot Program.

\appendix 
\begin{widetext}

\section{Classical Range-Delay Estimation with Random Phase}
\label{app:phase_noise}
To enable explicit comparisons between phase-coherent and phase-incoherent classical radars, we will consider a classical radar that uses ideal heterodyne reception instead of optimum quantum reception, as analysis of phase-incoherent operation of the latter poses major challenges.  Nevertheless, in the phase-coherent case with high-brightness ($N_B \gg 1$) background radiation---as will be true for microwave operation---a classical radar using heterodyne reception is known to approximate optimum quantum reception for that radar.  We expect the same is likely to prevail for phase-incoherent operation.  

The complex field envelope of the intermediate frequency signal-plus-noise waveform for a heterodyne reception classical radar can be taken to 
be~\cite{Shapiro2009}
\be
{\bf E}_R(t) = \sqrt{\kappa \calE}\,{\bf s}(t-\tau)e^{i\theta_R} + {\bf w}(t),
\label{Er}
\ee
where: $\kappa$ is the roundtrip radar-to-target-to-radar transmissivity of the target return; $\tau$ is the target's range delay; ${\bf s}(t)$ is the transmitter's normalized ($\int\!{\rm d}t\,|{\bf s}(t)|^2 = 1$) duration $T$, bandwidth $\Delta\omega$ pulse shape; $\theta_R$ is the phase shift incurred in reflection from the target; and ${\bf w}(t)$ is a zero-mean, complex-valued, white Gaussian noise process characterized by its correlation function, $\langle {\bf w}^*(t){\bf w}(t')\rangle = (N_B + 1)\delta(t-t')$, with $N_B \gg 1$ being the background radiation's brightness and $\delta(\cdot)$ the unit impulse.
(The main text took ${\bf s}(t)$ to be a high time-bandwidth product, chirped-Gaussian pulse, but for most of this section we shall allow it to be an arbitrary waveform satisfying the constraints mentioned above.)

The preceding radar makes its range-delay estimate using $\{{\bf E}_R(t) : t\in \mathcal{T}\}$, where $\mathcal{T}$ includes all times for which there could be a target return from the range-delay uncertainty interval $\tau \in [\tau_{\rm min},\tau_{\rm max}]$.  For now, let us assume that $\mathcal{T}$ has finite duration $T_R$.  Exploiting the whiteness of ${\bf w}(t)$, the continuous-time observation $\{{\bf E}_R(t) : t\in \mathcal{T}\}$ can be discretized using its Fourier series coefficients, $\{{\bf r}_n : |n| = 0,1,2,\ldots\}$, given by
\be
{\bf r}_n \equiv \int_{\mathcal{T}}\!{\rm d}t\,{\bf E}_R(t)\frac{\displaystyle e^{-i2\pi nt/T_R}}{\displaystyle \sqrt{T_R}},
\ee
with
\begin{eqnarray}
{\bf r}_n &=& \sqrt{\kappa\cal}\,{\bf s}_n(\tau)e^{i\theta_R} + {\bf w}_n,\\[.05in]
{\bf s}_n(\tau) &\equiv& \int_{\mathcal{T}}\!{\rm d}t\,{\bf s}(t-\tau)\frac{\displaystyle e^{-i2\pi nt/T_R}}{\displaystyle \sqrt{T_R}},\\[.05in]
{\bf w}_n &\equiv&  \int_{\mathcal{T}}\!{\rm d}t\,{\bf w}(t)\frac{\displaystyle e^{-i2\pi nt/T_R}}{\displaystyle \sqrt{T_R}},
\end{eqnarray}
and the $\{{\bf w}_n\}$ being independent identically-distributed (iid), zero-mean, complex-valued Gaussian random variables with $\langle |{\bf w}_n|^2\rangle = N_B + 1.$
This radar's Cram\'{e}r-Rao bound (CRB) and the Ziv-Zakai bound (ZZB) are treated in the subsections that follow.  Both analyses start from the conditional likelihood function,
\be
p_{\{{\bf r}_n\}\mid \theta_R}(\{{\bf R}_n\}\mid \Theta_R) = \prod_n
\frac{\displaystyle \exp[-|{\bf R}_n-\sqrt{\kappa\calE}\,{\bf s}_n(\tau)e^{i\Theta_R}|^2/(N_B+1)]}{\displaystyle \pi(N_B+1)}.
\label{classicalLikelihood}
\ee 
The classical radar's phase-coherent CRB and ZZB from the main text follow from assuming $\Theta_R = 0$ in \eqref{classicalLikelihood}.  These will be reprised below as  preludes to our results for phase-incoherent ($\theta_R$ uniformly distributed on $[0,2\pi]$) operation.

\subsection{Cram\'{e}r-Rao Bound}
\label{app:phase_noiseCRB}
The main text's phase-coherent CRB for the classical radar's mean-squared range-delay accuracy is the reciprocal of its Fisher information,
\be
\calF_{\rm coh} = \left\langle\!\left(\frac{\displaystyle \partial \ln[p_{\{{\bf r}_n\}\mid \theta_R}(\{{\bf r}_n\}\mid 0) ]}{\displaystyle \partial \tau}\right)^2\right\rangle,
\ee
which evaluates to 
\ba
\calF_{\rm coh} &= \frac{\displaystyle \left\langle \!\left(2\sqrt{\kappa\calE}\,\sum_n{\rm Re}\!\left\{[{\bf r}_n-\sqrt{\kappa \calE}\,{\bf s}_n(\tau)]\frac{{\rm d}{\bf s}^*_n(\tau)}{{\rm d}\tau}\right\}\right)^2\right\rangle}{\displaystyle (N_B + 1)^2}\\[.05in]
&= \frac{\displaystyle 2\kappa\calE \sum_n \left|\frac{{\rm d}{\bf s}_n(\tau)}{{\rm d}\tau}\right|^2}{\displaystyle N_B+1} = \frac{\displaystyle 2\kappa \calE\int_{\mathcal{T}}\!{\rm d}t\,\left|\frac{{\rm d}{\bf s}(t-\tau)}{{\rm d}\tau}\right|^2}{\displaystyle N_B+1} = \frac{\displaystyle 2\kappa\calE \Delta\omega^2}{\displaystyle N_B+1},
\label{phasecohCRB}
\ea
where the second and third equalities in Eq.~\eqref{phasecohCRB} follow, respectively,  from Parseval's theorem for Fourier series and $\Delta\omega$ defined to be the root-mean-squared (rms) bandwidth of ${\bf s}(t)$.  At this point we can let $T_r \rightarrow \infty$, and, using $N_B \gg 1$ and signal-to-noise ratio (SNR) defined by ${\rm SNR} \equiv \kappa\calE/N_B$, we get the main text's Eq.~\eqref{dt_C} for an arbitrary ${\bf s}(t)$ satisfying the assumed constraints. 

Turning now to phase-incoherent operation, we start from the likelihood function
\begin{eqnarray}
p_{\{{\bf r}_n\}}(\{{\bf R}_n\}) &=& \int_0^{2\pi}\!{\rm d}\Theta_R\,\frac{p_{\{{\bf r}_n\}\mid \theta_R}(\{{\bf R}_n\}\mid \Theta_R)}{2\pi} \\[.05in]
&=& \frac{\displaystyle \exp\!\!\left[-\frac{\sum_n(|{\bf R}_n|^2+ \kappa\calE|{\bf s}_n(\tau)|^2 )}{N_B+1}\right]I_0\!\!\left[\frac{2\sqrt{\kappa\calE}\,|\sum_n{\bf R}_n{\bf s}_n^*(\tau)|}{N_B+1}\right]}{\displaystyle \prod_n\pi(N_B+1)},
\end{eqnarray}
where $I_0(\cdot)$ is the zeroth-order modified Bessel function of the first kind.  Taking the logarithm and discarding terms that will disappear after differentiation by $\tau$, we get
\be
\mathcal{F}_{\rm incoh} = \left\langle\!\left[\frac{\partial}{\partial\tau}\left(-\kappa\calE\sum_n|{\bf s}_n(\tau)|^2/(N_B+1) + \ln\!\left\{I_0\!\!\left[\frac{2\sqrt{\kappa\calE}\,|\sum_n{\bf r}_n{\bf s}_n^*(\tau)|}{N_B+1}\right]\right\}\right)\right]^2\right\rangle.
\ee
Parseval's theorem for Fourier series now gives us
\be
\mathcal{F}_{\rm incoh} = \left\langle\!\left[\frac{\partial}{\partial\tau}\left(-\kappa\calE\int_{\mathcal{T}}\!{\rm d}t\,|{\bf s}(t-\tau)|^2/(N_B+1) + \ln\!\left\{I_0\!\!\left[\frac{2\sqrt{\kappa\calE}\,|\int_{\mathcal{T}}{\rm d}t\,{\bf r}(t){\bf s}^*(t-\tau)|}{N_B+1}\right]\right\}\right)\right]^2\right\rangle,
\ee
which reduces to
\be
\mathcal{F}_{\rm incoh} = \left\langle\!\left( \frac{\displaystyle I_1\!\!\left[\frac{2\sqrt{\kappa\calE}\,|\int_{\mathcal{T}}{\rm d}t\,{\bf r}(t){\bf s}^*(t-\tau)|}{N_B+1}\right]}
{\displaystyle I_0\!\!\left[\frac{2\sqrt{\kappa\calE}\,|\int_{\mathcal{T}}{\rm d}t\,{\bf r}(t){\bf s}^*(t-\tau)|}{N_B+1}\right]}\frac{2\sqrt{\kappa\calE}}{N_B+1}\frac{\partial |\int_{\mathcal{T}}{\rm d}t\,{\bf r}(t){\bf s}^*(t-\tau)|}{\partial\tau}
\right)^2\right\rangle,
\ee
by virtue of $\int_{\mathcal{T}}{\rm d}t\,|{\bf s}(t-\tau)|^2 = 1$ and ${\rm d}I_0(x)/{\rm d}x = I_1(x)$, where $I_1(\cdot)$ is the first-order modified Bessel function of the first kind.

At this point, we must choose a specific ${\bf s}(t)$ and switch to numerical evaluation for our comparison between the phase-coherent and phase-incoherent CRBs.  To ease the ensuing numerical burden, we assume the transform-limited Gaussian pulse ${\bf s}(t) = e^{-t^2/4T^2}/(2\pi T^2)^{1/4}$, whose mean-squared bandwidth is $\Delta\omega^2 = 1/4T^2$.  Next, we use the discretization
\be
\int_{\mathcal{T}}{\rm d}t\,{\bf r}(t){\bf s}^*(t-\tau) \approx \sum_{m\in \mathcal{M}}\tilde{\bf r}_m\tilde{\bf s}_m(\tau),
\ee
where 
\be
\tilde{\bf r}_m \equiv \int_{(m-1/2)\Delta t}^{(m+1/2)\Delta t}\!{\rm d}t\,\frac{{\bf r}(t)}{\sqrt{\Delta t}},
\ee
and
\be
\tilde{\bf s}_m(\tau) \equiv \int_{(m-1/2)\Delta t}^{(m+1/2)\Delta t}\!{\rm d}t\,\frac{{\bf s}(t-\tau)}{\sqrt{\Delta t}},
\ee
with $\Delta t$ being small enough that $\tilde{\bf s}_m(\tau) \approx {\bf s}(m\Delta t-\tau)$, ${\rm d}\tilde{\bf s}_m(\tau)/{\rm d}\tau \approx {\rm d}{\bf s}(m\Delta t-\tau)/{\rm d}\tau$, and $m\in\mathcal{M}$ such that $t\in [\min_m(m\Delta t),\max_m(m\Delta t)]$ spans $\mathcal{T}$.  Given $\theta_R = \Theta_R$, the $\{\tilde{\bf r}_m\}$ are statistically independent, complex-valued, Gaussian random variables with conditional means 
\be 
\mathbb{E}(\tilde{\bf r}_m\mid \theta_R= \Theta_R) = \sqrt{\kappa\calE}\,\tilde{\bf s}_m(\tau)e^{i\Theta_R},
\ee
and conditional variances 
\be
{\rm Var}(\tilde{\bf r}_m\mid \theta_R= \Theta_R) = \langle |\tilde{\bf w}_m|^2\rangle = N_B + 1,
\ee
where
\be
\tilde{\bf w}_m \equiv \int_{(m-1/2)\Delta t}^{(m+1/2)\Delta t}\!{\rm d}t\,\frac{{\bf w}(t)}{\sqrt{\Delta t}}, \mbox{ for $m \in \mathcal{M}$}
\ee
is an iid collection of zero-mean, complex-Gaussian random variables, i.e., a discrete-index, complex-valued, white Gaussian noise. 

\begin{figure}[h]
\vspace{5 mm}
    \centering
    \includegraphics[width=0.475\textwidth]{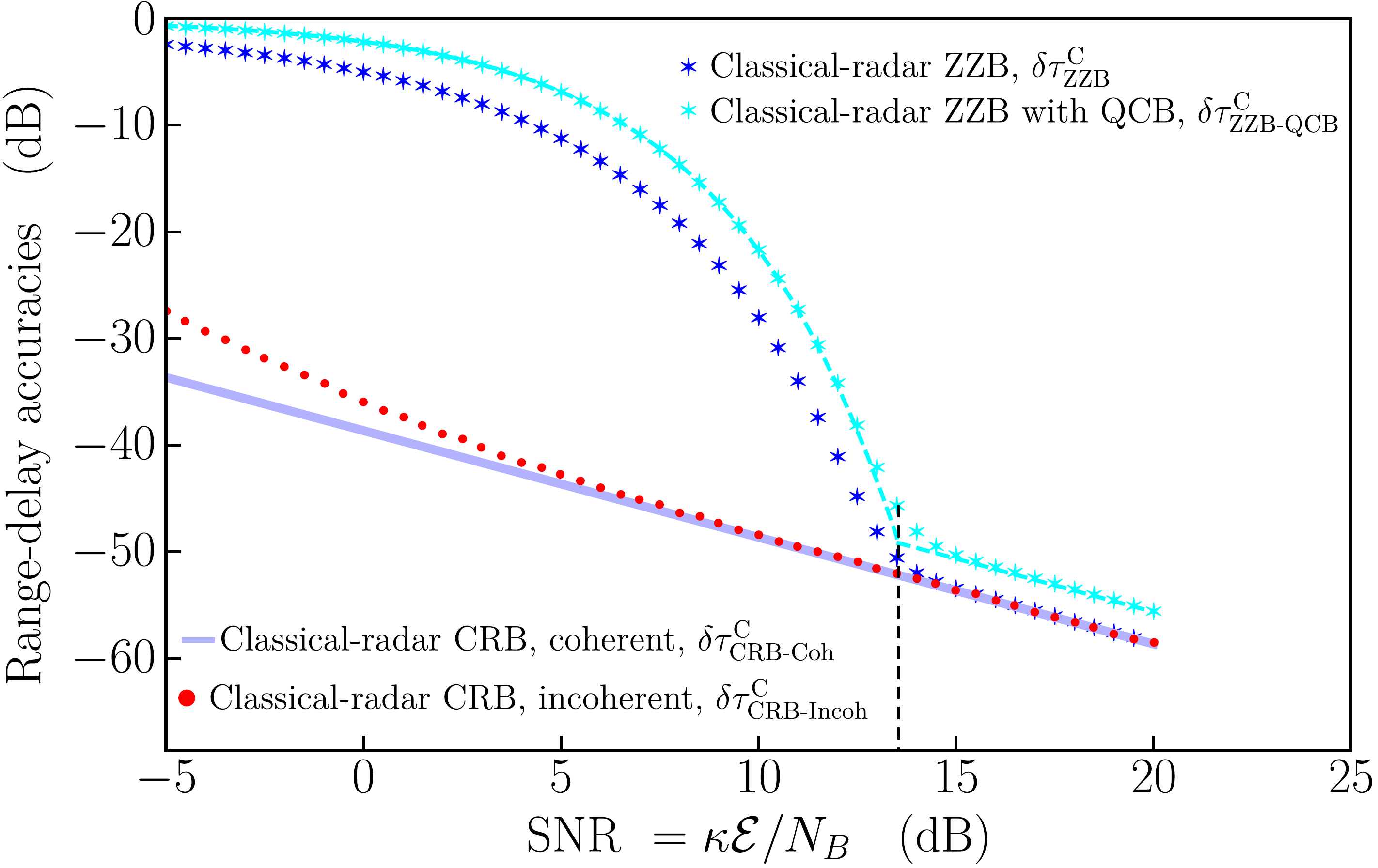}
   \caption{Normalized mean-squared range-delay accuracies in dB,  $20\log_{10}(\delta\tau/\sigma_\tau)$, versus SNR in dB.  The plots assume $\Delta\omega/2\pi = 10^6\,$Hz and $\Delta R = 5$\,km.  Results are shown, top to bottom, for $\delta\tau_{\rm ZZB\text{-}QCB}^{\rm C}$ (cyan stars), $\delta\tau_{\rm ZZB}^{\rm C}$ (blue stars), $\delta\tau_{\rm CRB\text{-}Incoh}^{\rm C}$ (red dots), and $\delta\tau_{\rm CRB\text{-}Coh}^{\rm C}$ (blue line).  Also plotted is a vertical dashed line showing the classical radar's SNR threshold, ${\rm SNR}_{\rm thresh}^{\rm C}$.  See the main text for information about the ZZBs and the threshold SNR.
    \label{fig:radar_scenario_phase_noise}
}
\end{figure}
It is now straightforward to evaluate the phase-incoherent Fisher information by Monte Carlo integration and thus obtain the CRB on the rms range-delay accuracy, $\delta\tau_{\rm CRB\text{-}{\rm Incoh}}^{\rm C} = 1/\sqrt{\mathcal{F}_{\rm incoh}}$, for comparison with its phase-coherent counterpart, $\delta\tau_{\rm CRB\text{-}{\rm Coh}}^{\rm C}$, given by the main text's Eq.~\eqref{dt_C}.     Figure~\ref{fig:radar_scenario_phase_noise} shows the results of doing so for $\Delta\omega/2\pi = 10^6$\,Hz, i.e., the same rms bandwidth used in the main text's Fig.~\ref{fig:dR}. It contains dB plots of the preceding range-delay accuracies normalized by the rms range-delay error, $\sigma_\tau \equiv \Delta\tau/\sqrt{12}$, of a uniform distribution over a $\Delta\tau \equiv 2\Delta R/c$ duration range-delay uncertainty interval, where $\Delta R = 5\,$km and $c$ is light speed.  From this figure we see that $\delta\tau_{\rm CRB\text{-}{\rm Incoh}}^{\rm C}$ converges to $\delta\tau_{\rm CRB\text{-}{\rm Coh}}^{\rm C}$ with increasing SNR.  Furthermore, and more importantly, because Fig.~\ref{fig:radar_scenario_phase_noise} also includes the ZZB results $\delta\tau_{\rm ZZB}^{\rm C}$ and $\delta\tau_{\rm ZZB\text{-}QCB}^{\rm C}$ from the main text's Fig.~\ref{fig:dR}, we see that the above-threshold CRBs for phase-coherent and phase-incoherent operation coincide where it matters, i.e., where rms range-delay accuracy is given by the CRB rather than just bounded by it.

\subsection{Ziv-Zakai Bound}
Our classical radar's Ziv-Zakai bound is
\begin{align}
\delta \tau_{\rm ZZB}=
\sqrt{\int_0^{\Delta\tau}\!{\rm d}\tau'\, \tau'\!\left(1-\frac{\tau'}{\Delta\tau}\right)\!P_e(\tau')},
\label{Ziv_Zakai_classical}
\end{align}
where $P_e(\tau')$ is the minimum error probability for deciding between the equally-likely hypotheses $H_0$ = target present at range delay $\tau_{\rm min}$ and $H_1$ = target present at range delay $\tau_{\rm min}+\tau'$ based on observation of $\{{\bf E}_R(t) : t\in \mathcal{T}\}$.  For the phase-coherent ($\theta_R = 0$) case treated in the main text, the decision rule that gives the minimum error-probability decision comes from a likelihood-ratio test (LRT)~\cite{VanTrees2001a} that simplifies to
\be
\int_{\mathcal{T}}\!{\rm d}t\,{\bf r}(t){\bf s}^*(t-\tau_{\rm min}-\tau')
\begin{array}{c}
\mbox{\scriptsize decide $H_1$}\\
\ge \\ 
< \\
\mbox{\scriptsize decide $H_0$}
\end{array}
\int_{\mathcal{T}}\!{\rm d}t\,{\bf r}(t){\bf s}^*(t-\tau_{\rm min}).
\label{LRTsimple}
\ee
From this result it is easily shown~\cite{VanTrees2001a} that
\ba
P_e(\tau') &= Q\!\left(\sqrt{\kappa\calE\!\int_{\mathcal{T}}{\rm d}t\,|{\bf s}(t-\tau_{\rm min})-{\bf s}^*(t-\tau_{\rm min}-\tau')|^2/2(N_B+1)}\right)
\label{cohPeGen}\\[.05in]
&\le \exp\!\left[-\kappa\calE\!\int_{\mathcal{T}}{\rm d}t\,|{\bf s}(t-\tau_{\rm min})-{\bf s}^*(t-\tau_{\rm min}-\tau')|^2/4(N_B+1)\right]/2,
\label{cohPeBdGen}
\ea
where $Q(x)\equiv \int_x^\infty\!{\rm d}y\,e^{-y^2/2}/\sqrt{2\pi}$ and the upper bound is both the well-known $Q(|x|) \le e^{-x^2/2}/2$ inequality and the Chernoff bound on $P_e(\tau')$.  

To get a simple, but insightful, result for the phase-incoherent case, we shall assume the radar uses the rectangular pulse shape~\cite{footnote1} 
\be
{\bf s}(t) = \left\{\begin{array}{ll}
1/\sqrt{T_s}, & \mbox{for $0 \le t\le T_s$,}\\[.05in]
0, & \mbox{otherwise.}
\end{array}\right.
\label{rectpulse}
\ee
For phase-coherent operation, Eq.~\eqref{cohPeGen} becomes
\be
P_e(\tau') = \left\{\begin{array}{ll}
Q\!\left[\sqrt{\kappa\calE\tau'/T_s(N_B+1)}\right], &\mbox{for $0 \le \tau'\le T_s$},\\[.1in]
Q\!\left[\sqrt{\kappa\calE/(N_B+1)}\right], & \mbox{for $\tau' > T_s$,} \end{array}\right.
\ee
and the upper bound from \eqref{cohPeBdGen} becomes
\be
P_e(\tau') \le \left\{\begin{array}{ll}
\exp[-\kappa\calE\tau'/2T_s(N_B+1)]/2, &\mbox{for $0 \le \tau'\le T_s$},\\[.1in]
\exp[-\kappa\calE/2(N_B+1)]/2, & \mbox{for $\tau' > T_s$.} \end{array}\right.
\label{cohPeBdRect}
\ee

To treat phase-incoherent operation with ${\bf s}(t)$ from Eq.~\eqref{rectpulse}, we note that for the just considered $\theta_R=0$ case, 
\begin{equation}
\br_0 \equiv \frac{1}{\sqrt{\tau'}} \int_{\tau_{\rm min}}^{\tau_{\rm min} + \tau'}\!{\rm d}t\,\br(t),
\end{equation}
\begin{equation}
\br \equiv \frac{1}{\sqrt{T_s-\tau'}}\int_{\tau_{\rm min}+\tau'}^{\tau_{\rm min}+T_s}\!{\rm d}t\,\br(t),
\end{equation}
and
\begin{equation}
\br_1 \equiv \frac{1}{\sqrt{\tau'}}\int_{\tau_{\rm min}+T_s}^{\tau_{\rm min}+\tau' + T_s}\!{\rm d}t\,\br(t),
\end{equation}
are sufficient statistics for the minimum error-probability test in \eqref{LRTsimple} when $0\le \tau'\le T_s$, as are
\begin{equation}
\br'_0 \equiv \frac{1}{\sqrt{T_s}} \int_{\tau_{\rm min}}^{\tau_{\rm min} + T_s}\!{\rm d}t\,\br(t),
\end{equation}
and
\begin{equation}
\br'_1 \equiv \frac{1}{\sqrt{T_s}}\int_{\tau_{\rm min}+\tau'}^{\tau_{\rm min} + \tau'+T_s}\!{\rm d}t\,\br(t),
\end{equation}
when $\tau' > T_s$.  So, for phase-incoherent operation, let us assume that the radar decides between target absence or presence from LRTs based on $\{r_0 \equiv |\br_0|,r \equiv |\br|,r_1 \equiv |\br_1|\}$ for $0\le \tau'\le T_s$ and based on $\{r_0' \equiv |\br'_0|,r_1'\equiv |\br'_1|\}$ for $\tau' > T_s$.  

With $\theta_R$ uniformly distributed on $[0,2\pi]$ and $0\le \tau\le T_S$, the likelihoods for the two hypotheses are easily shown to be 
\begin{eqnarray}
p_{r_0,r,r_1\mid H_1}(R_0,R,R_1\mid H_1)  &=& 
\left[\frac{2R_0}{N_B+1}e^{-R_0^2/(N_B+1)}\right] \nonumber \\[.05in] &\times& \left[\frac{2R}{N_B+1}e^{-[R^2 + \kappa \calE (1-\tau'/T_s)]/(N_B+1)}I_0[2\sqrt{\kappa \calE (1-\tau'/T_s)}R/(N_B+1)]\right] \nonumber \\[.05in]
&\times&
\left[\frac{2R_1}{N_B+1}e^{-(R_1^2 + \kappa \calE \tau'/T_s)/(N_B+1)}I_0[2\sqrt{\kappa \calE \tau'/T_s}R_1/(N_B+1)]\right]\!,
\label{LFp1}
\end{eqnarray}
and
\begin{eqnarray}
p_{r_0,r,r_1\mid H_0}(R_0,R,R_1\mid H_0)  &=& 
\left[\frac{2R_0}{N_B+1}e^{-(R_0^2 + \kappa \calE \tau'/T_s)/N_0}I_0[2\sqrt{\kappa \calE \tau'/T_s}R_0/(N_B+1)]\right]
\nonumber \\[.05in] &\times& \left[\frac{2R}{N_B+1}e^{-[R^2 + \kappa \calE (1-\tau'/T_s)]/(N_B+1)}I_0[2\sqrt{\kappa \calE (1-\tau'/T_s)}R/(N_B+1)]\right] \nonumber\\[.05in]
&\times&
\left[\frac{2R_1}{N_B+1}e^{-R_1^2/(N_B+1)}\right].
\label{LFp0}
\end{eqnarray}  
The phase-incoherent LRT then reduces to 
\begin{equation}
I_0[2\sqrt{\kappa \calE \tau'/T_s}\,r_1/(N_B+1)] 
\begin{array}{c}
\mbox{\scriptsize decide $H_1$}\\
\ge \\ 
< \\
\mbox{\scriptsize decide $H_0$}
\end{array}
I_0[2\sqrt{\kappa \calE \tau'/T_s}\,r_0/(N_B+1)],
\end{equation}
which further reduces to the simple threshold test
\begin{equation}
r_1 \begin{array}{c}
\mbox{\scriptsize decide $H_1$}\\
\ge \\ 
< \\
\mbox{\scriptsize decide $H_0$}
\end{array}
r_0,
\end{equation}
because $I_0(|x|)$ is monotonically increasing with increasing $|x|$.  A very similar calculation will show that the phase-incoherent LRT for $\tau > T_s$ reduces to the threshold test
\begin{equation}
r'_1 \begin{array}{c}
\mbox{\scriptsize decide $H_1$}\\
\ge \\ 
< \\
\mbox{\scriptsize decide $H_0$}
\end{array}
r'_0.
\end{equation}

To find $P_e(\tau')$, it is easier to begin with the $\tau' >  T_s$ case.   There, we have that the false-alarm and miss probabilities satisfy
\begin{align}
P_F &\equiv \Pr(\mbox{decide $H_1$ when $H_0$ true}) = \Pr(r_1\ge r_0\mid H_0)\\[.05in]
&= \int_0^\infty\!{\rm d}R_0\,\frac{2R_0}{N_B+1}e^{-(R_0^2 + \kappa \calE)/(N_B+1)}I_0[2\sqrt{\kappa \calE}R_0/(N_B+1)]\int_{R_0}^\infty\!{\rm d}R_1\,\frac{2R_1}{N_B+1}e^{-R_1^2/(N_B+1)}\\[.05in]
&= \int_0^\infty\!{\rm d}R_0\,\frac{2R_0}{N_B+1}e^{-(2R_0^2 + \kappa \calE)/(N_B+1)}I_0[2\sqrt{\kappa E}R_0/(N_B+1)] = \frac{e^{-\kappa \calE/2(N_B+1)}}{2},
\end{align}
and
\begin{align}
P_M &\equiv \Pr(\mbox{decide $H_0$ when $H_1$ true}) = \Pr(r_0 > r_1\mid H_1)\\[.05in]
&= \int_0^\infty\!{\rm d}R_1\,\frac{2R_1}{N_B+1}e^{-(R_1^2 + \kappa \calE)/(N_B+1)}I_0[2\sqrt{\kappa \calE}R_1/(N_B+1)]\int_{R_1}^\infty\!{\rm d}R_0\,\frac{2R_0}{N_B+1}e^{-R_0^2/(N_B+1)}\\[.05in]
&= \int_0^\infty\!{\rm d}R_1\,\frac{2R_1}{N_B+1}e^{-(2R_1^2 + \kappa \calE)/(N_B+1)}I_0[2\sqrt{\kappa \calE}R_1/(N_B+1)] = \frac{e^{-\kappa \calE/2(N_B+1)}}{2}.
\end{align}
Because the two hypotheses are equally likely, we get
\be
P_e(\tau') = P_F/2 + P_M/2 = \frac{e^{-\kappa \calE/2(N_B+1)}}{2}, \mbox{ for $\tau' > T_s$}.
\label{PeIncohHiTau}
\ee
Paralleling what we did to find $P_e(\tau')$ for $\tau' >  T_s$, we find that 
\begin{equation}
P_e(\tau')= \frac{e^{-\kappa \calE\tau'/2T_s(N_B+1)}}{2}, \mbox{ for $0\le \tau' \le T_s$}.
\label{PeIncohLoTau}
\end{equation}
Note that these results for phase-incoherent operation \emph{exactly} match the $P_e(\tau')$ upper bounds for phase-coherent operation from \eqref{cohPeBdRect}.  Consequently, when $N_B \gg 1$, so that the Chernoff bound result in \eqref{cohPeBdRect} matches our classical radar's quantum Chernoff bound (QCB), we have that the phase-incoherent $\delta\tau_{\rm ZZB}^{\rm C}$ equals the phase-coherent $\delta\tau_{\rm ZZB\text{-}QCB}^{\rm C}$ for the rectangular pulse shape we have assumed herein. 

\section{Quantum Fisher Information}
\label{app:fisher}
In this section we derive the quantum Fisher informations for the main text's coherent-state and quantum pulse-compression radars.  For both radars we assume phase-coherent operation with the carrier-frequency phase shift suppressed.  Thus the Fourier-domain representation of the radars' received field operator, from the main text's Eq.~\eqref{phase_coherent}, is 
\be 
\hat{A}_R(\omega)=\sqrt{\kappa}\,e^{i\omega \tau} \hat{A}_S(\omega)+\sqrt{1-\kappa}\,\hat{A}_B(\omega).
\label{phase_coherent_SI}
\ee 
The radars' range-delay estimates are then made from a measurement on the received field operator
\be
\hat{E}_R(t) = \int\!\frac{{\rm d}\omega}{2\pi}\,\hat{A}_R(\omega)e^{-i\omega t} = \sqrt{\kappa}\,\hat{E}_S(t-\tau) + \sqrt{1-\kappa}\,\hat{E}_B(t),
\label{ER_app}
\ee
for $t\in \mathcal{T}$, where, as before, $\mathcal{T}$ includes all times for which there could be any target return from the range uncertainty interval.
  
\subsection{Quantum Fisher Information for the Coherent-State Radar}
To obtain the coherent-state radar's quantum Fisher information, we start from the Fourier-series mode decomposition,
\be
\hat{E}_R(t) = \sum_n \hat{a}_{R_n}\frac{e^{-i2\pi nt/T_R}}{\sqrt{T_R}},\mbox{ for $t\in \mathcal{T}$},
\ee
where $T_R$, the duration of $\mathcal{T}$, is assumed, for now, to be finite. It then follows that this decomposition's modal annihilation operators satisfy
\be 
\hat{a}_{R_n} = \sqrt{\kappa}\,\hat{a}_{S_n}e^{i2\pi n\tau/T_R} + \sqrt{1-\kappa}\,\hat{a}_{B_n},
\ee
where the modal annihilation operators, $\{\hat{a}_{S_n}\}$ and $\{\hat{a}_{B_n}\}$, for the transmitted signal and the background radiation are given by
\be
\hat{a}_{S_n} = \int_{\mathcal{T}}\!{\rm d}t\,\hat{E}_S(t-\tau)\frac{e^{-i2\pi n(t-\tau)/T_R}}{\sqrt{T_R}},
\ee
and
\be
\hat{a}_{B_n} = \int_{\mathcal{T}}\!{\rm d}t\,\hat{E}_B(t)\frac{e^{-i2\pi nt/T_R}}{\sqrt{T_R}}.
\ee
For the coherent-state radar whose transmitter emits $\hat{E}_S(t)$ in the coherent state $|\sqrt{\calE}\,{\bf s}(t)\rangle$~\cite{footnote2}, where $\int\!{\rm d}t\,|{\bf s}(t)|^2 = 1$, the $\hat{a}_{S_n}$ mode will be in the coherent state $|S(2\pi n/T_R)/\sqrt{T_R}\rangle$, where $S(\omega) \equiv \int\!{\rm d}t\,{\bf s}(t)e^{-i\omega t}$.  The $\{\hat{a}_{B_n}\}$ modes are in iid thermal states with average photon number $N_B/(1-\kappa)$, where, as in the main text, we assume $N_B \gg 1$ and $\kappa \ll 1$.  

The Fisher information we are seeking is
\be
\mathcal{F}^{\rm C}_\tau = \lim_{{\rm d}\tau\rightarrow 0}
\frac{8\!\left(1-\sqrt{{\rm tr}\!\left[\!\left(\sqrt{\hat{\rho}_\tau}\,\hat{\rho}_{\tau+{\rm d}\tau}\sqrt{\hat{\rho}_\tau}\right)^2\right]}\right)}{{\rm d}\tau^2},
\label{FisherStart}
\ee
where $\hat{\rho}_u$ is the state of $\{\hat{E}_R(t) : t\in \mathcal{T}\}$ when the range delay is $u$. Using the preceding paragraph's mode decomposition, we have that
$\mathcal{F}^{\rm C}_\tau = \sum_n\mathcal{F}^{\rm C}_n,$ 
where $\mathcal{F}^{\rm C}_n$ is the Fisher information  about $\tau$ that is embedded in the $\hat{a}_{R_n}$ mode's target-return phase rotation, $\phi_n = 2\pi n\tau/T_R$.  Thus, if we find the Fisher information $\mathcal{F}^{\rm C}_{\phi_n}$ for $\hat{a}_{R_n} = \sqrt{\kappa}\,\hat{a}_{S_n}e^{i\phi_n} + \sqrt{1-\kappa}\,\hat{a}_{B_n}$, we can get $\mathcal{F}^{\rm C}_n$ via $\mathcal{F}^{\rm C}_n = (2\pi n/T_R)^2\mathcal{F}^{\rm C}_{\phi_n}$.

To complete our derivation we therefore need to evaluate
 \be
\mathcal{F}^{\rm C}_{\phi_n} = \lim_{{\rm d}\phi\rightarrow 0}
\frac{8\!\left(1-\sqrt{{\rm tr}\!\left[\!\left(\sqrt{\hat{\rho}_{\phi_n}}\,\hat{\rho}_{\phi_n+{\rm d}\phi}\sqrt{\hat{\rho}_{\phi_n}}\right)^2\right]}\right)}{{\rm d}\phi^2},
\ee
where $\hat{\rho}_{\phi_n}$ is the state of the $\hat{a}_{R_n}$ mode when its $\hat{a}_{S_n}$-mode component has phase shift $\phi_n$.   The desired $\mathcal{F}^{\rm C}_{\phi_n}$ result is easily obtained~\cite{scutaru1998fidelity}:
\be 
\calF^{\rm C}_{\phi_n}=\frac{2\kappa\calE |S(2\pi n/T_R)|^2}{T_R(N_B+1/2)}.
\label{J_coh}
\ee 
Using this result gives us 
\be
\calF^{\rm C}_{\tau} = \sum_n\calF^{\rm C}_n = \sum_n\left(\frac{2\pi n}{T_R}\right)^2 \frac{2\kappa\calE |S(2\pi n/T_R)|^2}{T_R(N_B+1/2)}.
\label{FcohPenult} 
\ee
Finally, passing to the limit $T_R \rightarrow \infty$, we obtain
\be
\calF^{\rm C}_{\tau} =  \int\!\frac{{\rm d}\omega}{2\pi}\,\frac{2\kappa\calE \omega^2 |S(\omega)|^2}{N_B+1/2} = \frac{2\kappa\calE \Delta\omega^2}{N_B+1/2},
\label{FcohUlt}
\ee
where $\Delta\omega^2 = \int\!\frac{{\rm d}\omega}{2\pi}\,\omega^2|S(\omega)|^2$ is the transmitted pulse's mean-squared bandwidth.  Equation~\eqref{FcohUlt} agrees with the first equality in the main text's Eq.~\eqref{dt_C} and, using $N_B \gg 1$, it becomes $\mathcal{F}^{\rm C}_\tau = 2\Delta\omega^2{\rm SNR}$ as shown in the second equality in the main text's Eq.~\eqref{dt_C}.  The latter result also matches what was found here in Sec.~\ref{app:phase_noiseCRB} for the classical (i.e., coherent-state) radar that uses heterodyne reception.  

Note that our $\calF^{\rm C}_\tau$ derivation for the classical radar applies for all narrowband pulse shapes, i.e., it is \emph{not} restricted to the high time-bandwidth product, chirped Gaussian pulse that was employed in the main text.  Thus, in particular, it demonstrates that \emph{all} coherent-state radars of the same rms bandwidth and transmitted energy have the same quantum CRB for range-delay estimation in the $\kappa \ll 1$, $N_B \gg 1$ regime.

\subsection{Quantum Fisher Information for the Quantum Pulse-Compression Radar}
\label{QFIsection}
Our derivation of the quantum pulse-compression's quantum Fisher information, $
\calF^{\rm Q}_\tau$, is slightly more complicated than what sufficed for the coherent-state radar.  Now, the range-delay estimate should be based on joint measurement on $\{\hat{E}_R(t): t\in \mathcal{T}\}$ and $\{\hat{E}_I(t): t \in [0,T]\}$; The return field is determined by the signal field $\hat{E}_S(t)$ via Eq.~\eqref{ER_app}, where $\hat{E}_S(t)$ and $\hat{E}_I(t)$ are the signal and idler beams produced by a continuous-wave, frequency-degenerate, spontaneous parametric downconverter.  To obtain a useful mode decomposition for obtaining $\mathcal{F}^{\rm Q}_\tau$, it suffices to limit the received field measurement to $t\in [\tau,\tau+T]$, because ${\rm d}\tau$ in Eq.~\eqref{FisherStart} will be driven to zero.  
The mode decompositions we will now use are thus as follows:
\begin{eqnarray}
\hat{E}_R(t) &=& \sum_n \hat{a}_{R_n}\frac{e^{-i2\pi nt/T}}{\sqrt{T}},\mbox{ for $t\in [\tau,\tau+T]$},\\[.05in]
\hat{E}_S(t-\tau) &=& \sum_n\hat{a}_{S_n}\frac{e^{-i2\pi n(t-\tau)/T}}{\sqrt{T}},\mbox{ for $t\in [\tau,\tau+T]$},\\[.05in]
\hat{E}_B(t) &=& \sum_n \hat{a}_{B_n}\frac{e^{-i2\pi nt/T}}{\sqrt{T}},\mbox{ for $t\in [\tau,\tau+T]$},\\[.05in]  
\hat{E}_I(t) &=& \sum_n\hat{a}_{I_n}\frac{e^{i2\pi nt/T}}{\sqrt{T}},\mbox{ for $t\in [0,T]$}.
\end{eqnarray}
Then, because the quantum radar's transmitted signal and retained idler have high time-bandwidth products, $T\gg 2\pi/\Delta \omega$, the preceding Fourier series are the quantum version of Van Trees' stationary problem, long observation time (SPLOT) approximation to the Karhunen-Lo\'{e}ve expansion of the quantum radar's received and retained fields over their measurement intervals~\cite{VanTrees2001a}.  Consequently, $\{\hat{a}_{S_n},\hat{a}_{B_n},\hat{a}_{I_n}\}$ is a collection of statistically independent mode triplets.  Moreover, each triplet is in a zero-mean, jointly-Gaussian state with the $\hat{a}_{B_n}$ mode being in a thermal state with average photon number $N_B/(1-\kappa)$ that is statistically independent of the $\{\hat{a}_{S_n},\hat{a}_{I_n}\}$ mode pair.  That signal-idler mode pair is in a two-mode squeezed vacuum state with Wigner covariance matrix
\be
\Lambda_{S_nI_n} = \frac{1}{4}\!
\left[
\begin{array}{cc}
[2S^{(n)}(2\pi n/T) + 1] {\bf I}&2\sqrt{S^{(n)}(2\pi n/T)(S^{(n)}(2\pi n/T)+1)}\,{\bf R}_0\\[.05in]
2\sqrt{S^{(n)}(2\pi n/T)(S^{(n)}(2\pi n/T)+1)}\,{\bf R}_0&[2S^{(n)}(2\pi n/T)+1]{\bf I}
\end{array} 
\right],
\ee
where: $S^{(n)}(\omega)$ is the (signal and idler's) fluorescence spectrum from the main text's Eq.~\eqref{SnOmega}; ${\bf I}$ is the $2\times 2$ identity matrix; and ${\bf R}_\phi \equiv {\rm Re}\left[ \exp(i\phi)({\bf Z}-i{\bf X})\right]$, with ${\bf Z}$ and ${\bf X}$ being Pauli matrices.  From $\hat{a}_{R_n} = \sqrt{\kappa}\,\hat{a}_{S_n}e^{i\phi_n} + \sqrt{1-\kappa}\,\hat{a}_{B_n}$ with $\phi_n = 2\pi n\tau/T$ we then have that
$\{\hat{a}_{R_n},\hat{a}_{I_n}\}$ is a collection of independent mode pairs each of which is in a zero-mean, jointly-Gaussian state with Wigner covariance matrix
\be
\Lambda_{R_nI_n} = \frac{1}{4}\!
\left[
\begin{array}{cc}
[2\kappa S^{(n)}(2\pi n/T) +2N_B + 1] {\bf I}&2\sqrt{\kappa S^{(n)}(2\pi n/T)(S^{(n)}(2\pi n/T)+1)}\,{\bf R}_{\phi_n}\\[.05in]
2\sqrt{\kappa S^{(n)}(2\pi n/T)(S^{(n)}(2\pi n/T)+1)}\,{\bf R}_{\phi_n}&[2S^{(n)}(2\pi n/T)+1]{\bf I}
\end{array} 
\right].
\label{hk}
\ee 

At this point we can use methods from Ref.~\cite{marian2016quantum} to obtain 
\be 
\calF^{\rm Q}_{\phi_n}=\frac{4\kappa S^{(n)}(2\pi n/T)\left[S^{(n)}(2\pi n/T)+1\right]}{1+N_B\left[2S^{(n)}(2\pi n/T)+1\right]+\left(1-\kappa\right)S^{(n)}(2\pi n/T)}.
\label{J_TMSV}
\ee 
Paralleling what we did for the coherent-state radar, we can then find $\mathcal{F}^{\rm Q}_\tau$ via
\be
\mathcal{F}^{\rm Q}_\tau = \sum_n\left(\frac{2\pi n}{T}\right)^2 \calF^{\rm Q}_{\phi_n}.
\label{FtmsvPrelim}
\ee
Our interest is in the $S^{(n)}(\omega)\ll 1$, $\kappa\ll 1$, $N_B \gg 1$ regime, for which Eq.~\eqref{J_TMSV} reduces to $\mathcal{F}^{\rm Q}_{\phi_n} = 4\kappa S^{(n)}(2\pi n/T)/N_B$, so that Eq.~\eqref{FtmsvPrelim} yields
\be
\mathcal{F}^{\rm Q}_\tau = \sum_n\left(\frac{2\pi n}{T}\right)^2 \!\frac{4\kappa S^{(n)}(2\pi n/T)}{N_B} \approx T\int\!\frac{{\rm d}\omega}{2\pi}\,\frac{4\kappa \omega^2 S^{(n)}(\omega)}{N_B} = 4\Delta\omega^2{\rm SNR},
\label{tmsvFtau}
\ee
where the approximation's validity is due to having $T \gg 2\pi/\Delta\omega$.  Note that this $\calF^{\rm Q}_\tau$ result applies for all fluorescence spectra with rms bandwidth $\Delta\omega$. 

The quantum pulse-compression radar's $\calF^{\rm Q}_\tau$ we have just obtained deserves some discussion.  We have shown that  $\mathcal{F}_\tau^{\rm Q} = 2\mathcal{F}_\tau^{\rm C}$, as noted in the main text.  But does that $\mathcal{F}^{\rm Q}_\tau$ represent the ultimate performance limit for microwave radar's lossy ($\kappa \ll 1$), noisy ($N_B \gg 1$) operating regime?  It turns out that it does, as we now will show.

An upper bound on $\calF^{\rm Q}_\phi$ was recently obtained by Gagatsos~\emph{et al}.~\cite{gagatsos2017bounding} for estimating $\phi$ from a lossy (transmissivity $\kappa < 1$), noisy (average noise photon number per mode $N_B$) observation of a phase-shifted signal field. The signal field is in a state with mean photon number $N_S$ and photon-number variance $\Delta N_S^2$, and that state may be  multi-mode and entangled with an arbitrary ancilla.  Their bound's full expression is lengthy, but it is straightforward to show that its maximum, achieved in the limit $\Delta N_S^2 \rightarrow \infty$, is
\begin{align}
\calF^{\rm UB}_{\phi}=
\frac{4\kappa N_S\left(\kappa N_S+\left(1-\kappa\right)N_B+1\right)}{\left(1-\kappa\right)\!\left[\kappa N_S \left(2N_B+1\right)-\kappa N_B\left(N_B+1\right)+\left(N_B+1\right)^2\right]}.
\label{UB}
\end{align} 
For $\kappa\ll 1$, $\kappa N_S\ll N_B$, and  $N_B\gg 1$, we have
$\calF^{\rm UB}_\phi \rightarrow  4\kappa N_S/N_B$.  Applying this bound to the quantum radar's $\calF^{\rm Q}_{\phi_n}$ by identifying $N_S = S^{(n)}(2\pi n/T)$, we see that $\calF^{\rm Q}_{\phi_n}$ saturates the upper bound, implying that, for $\kappa \ll 1$ and $N_B \gg 1$, our quantum pulse-compression radar has the best high-SNR performance of all entanglement-assisted radars that use low-brightness transmitters.  

\section{Ziv-Zakai Bound  Evaluations}
\label{app:ziv-zakai}

In this section we assume phase-coherent operation and: evaluate the resulting ZZBs for our classical and quantum pulse-compression radars; derive their respective low-SNR and high-SNR asymptotic behaviors; and use those asymptotic behaviors to obtain expressions for their threshold SNRs, i.e., the SNRs below which their performance diverges from their CRBs.

\subsection{Classical Radar's Ziv-Zakai Bound}
\label{app:classical}

Our classical radar's description was given in Sec.~\ref{app:phase_noise}.  Its Ziv-Zakai bound was given in Eq.~\eqref{Ziv_Zakai_classical}, from which obtain the main text's $\delta\tau^{\rm C}_{\rm ZZB}$ by using the exact expression from Eq.~\eqref{cohPeGen} for $P_e(\tau')$.  The main text's $\delta\tau^{\rm C}_{\rm ZZB\text{-}QCB}$ is obtained by using  the quantum Chernoff bound (QCB) from \eqref{cohPeBdGen} in Eq.~\eqref{Ziv_Zakai_classical} in place of the exact error probability.  We shall leave the radar's pulse shape, ${\bf s}(t)$, arbitrary except for its being normalized, $\int\!{\rm d}t\,|{\bf s}(t)|^2 = 1$, and having Fourier transform $S(\omega)$ with mean-squared bandwidth $\int\!\frac{{\rm d}\omega}{2\pi}\,\omega^2|S(\omega)|^2 = \Delta\omega^2$.  Assuming $N_B \gg 1$, we then get
\begin{eqnarray}
P_e(\tau') &=& Q\!\left(\sqrt{{\rm SNR}\!\int\!\frac{{\rm d}\omega}{2\pi}\,\frac{|S(\omega)|^2|1-e^{-i\omega\tau'}|^2}{2}}\right)
\label{cohPeGenOmega}\\[.05in]
&\le& \frac{\displaystyle \exp\!\left(-{\rm SNR}\!\int\!\frac{{\rm d}\omega}{2\pi}\,\frac{|S(\omega)|^2|1-e^{-i\omega\tau'}|^2}{4})\right)}{\displaystyle 2}.
\label{cohPeBdGenOmega}
\end{eqnarray}
These expressions were used, with $S(\omega)/2\pi = e^{-\omega^2/2\Delta\omega^2}/\sqrt{2\pi\Delta\omega^2}$, to generate the $\delta\tau^{\rm C}_{\rm ZZB}$ and $\delta\tau^{\rm C}_{\rm ZZB\text{-}QCB}$ points shown in the main text's Fig.~\ref{fig:dR} and in Sec.~\ref{app:phase_noise}'s Fig.~\ref{fig:radar_scenario_phase_noise}.  All that remains, for our classical radar's ZZB, is to find its low-SNR and high-SNR asymptotic behaviors and the threshold SNR at which the former diverges from the latter.

At low SNRs, the principal contributions to $\delta\tau^{\rm C}_{\rm ZZB}$ and $\delta\tau^{\rm C}_{\rm ZZB\text{-}QCB}$ come from $\tau'$ in the vicinity of $\Delta\tau$.  Our radar scenario has $\Delta\tau \gg \tau_{\rm res} \sim 1/\Delta\omega$, so we can use 
\be
{\rm SNR}\!\int\!\frac{{\rm d}\omega}{2\pi}\,|S(\omega)|^2|1-e^{-i\omega\tau'}|^2 \approx
2\,{\rm SNR}\!\int\!\frac{{\rm d}\omega}{2\pi}\,|S(\omega)|^2 = 2\,{\rm SNR},
\ee
in Eq.~\eqref{cohPeGenOmega}, giving us the low-SNR asymptotic behaviors
\be
\delta\tau^{\rm C}_{\rm ZZB} \approx \sqrt{\int_0^{\Delta\tau}\!{\rm d}\tau'\, \tau'\!\left(1-\frac{\tau'}{\Delta\tau}\right)}\!Q\!\left(\sqrt{\rm SNR}\right) = \sqrt{\frac{\Delta\tau^2}{6}Q\!\left(\sqrt{\rm SNR}\right)},
\label{ClassLowSNR}
\ee
and
\be
\delta\tau^{\rm C}_{\rm ZZB\text{-}QCB} \approx \sqrt{\int_0^{\Delta\tau}\!{\rm d}\tau'\, \tau'\!\left(1-\frac{\tau'}{\Delta\tau}\right)\!\frac{e^{{-\rm SNR}/2}}{2}} = \sigma_\tau e^{-{\rm SNR}/4}.
\label{ClassQCBlowSNR}
\ee
Note that as ${\rm SNR}\rightarrow 0$, $\delta\tau^{\rm C}_{\rm ZZB}$ and 
$\delta\tau^{\rm C}_{\rm ZZB\text{-}QCB}$ both 
converge to the standard deviation, $\sigma_\tau$, of the range delay's prior distribution.  

At high SNRs the principal contributions to $\delta\tau^{\rm C}_{\rm ZZB}$ and $\delta\tau^{\rm C}_{\rm ZZB\text{-}QCB}$ come from $\tau'$ values near 0, so we can use
\be
{\rm SNR}\!\int\!\frac{{\rm d}\omega}{2\pi}\,|S(\omega)|^2|1-e^{-i\omega\tau'}|^2 \approx
{\rm SNR}\!\int\!\frac{{\rm d}\omega}{2\pi}\,|S(\omega)|^2\omega^2\tau'^2 = {\rm SNR}\,\Delta\omega^2\tau'^2,
\ee
in Eq.~\eqref{cohPeGenOmega} and in \eqref{cohPeBdGenOmega}, giving us the high-SNR asymptotic behaviors
\be
\delta\tau^{\rm C}_{\rm ZZB} \approx \sqrt{\int_0^{\Delta\tau}\!{\rm d}\tau'\,\tau'\!\left(1-\frac{\tau'}{\Delta\tau}\right)Q\!\left(\sqrt{\frac{{\rm SNR}\,\Delta\omega^2\tau'^2}{2}}\right)} \approx \frac{1}{\Delta\omega\sqrt{2\,{\rm SNR}}},
\label{ClassHiSNR}
\ee
and
\be
\delta\tau^{\rm C}_{\rm ZZB\text{-}QCB} \approx \sqrt{\int_0^{\Delta\tau}\!{\rm d}\tau'\, \tau'\!\left(1-\frac{\tau'}{\Delta\tau}\right)\frac{e^{-{\rm SNR}\,\Delta\omega^2\tau'^2/4}}{2}} \approx \frac{1}{\Delta\omega\sqrt{\rm SNR}}.
\label{ClassQCBhiSNR}
\ee
Equation~\eqref{ClassHiSNR} agrees with $\delta\tau^{\rm C}_{\rm CRB}$ from the main text's Eq.~\eqref{dt_C}, but Eq.~\eqref{ClassQCBhiSNR} is a factor of $\sqrt{2}$ higher, as discussed in the main text.

Our last ZZB task for the classical radar is to find ${\rm SNR}^{\rm C}_{\rm thresh}$. We do so by finding the SNR at which Eqs.~\eqref{ClassQCBlowSNR} and \eqref{ClassQCBhiSNR} agree.  The result we get is 
\be 
{\rm SNR}^{\rm C}_{\rm thresh} = 2 f\!\left(\frac{1}{2 \Delta \omega^2\sigma_\tau^2}\right),
\label{classThresh}
\ee
where $x=f(y)$ is the inverse function of $y=x e^{-x}$. 

\subsection{Quantum Pulse-Compression Radar's Ziv-Zakai Bound}
Our quantum pulse-compression radar was described in Sec.~\ref{QFIsection}.  As noted in the main text, we will use the quantum Chernoff bound on $P_e(\tau')$ in evaluating the main text's Eq.~\eqref{Ziv_Zakai_simplified}, because finding the quantum radar's exact error probability is too challenging.  Furthermore, we will restrict our attention to operation with $T \gg \Delta\tau$, because this condition is needed---for $S^{(n)}(\omega) \ll 1, \kappa\ll 1, N_B \gg 1$---to attain a useful SNR.  As a result, the mode decompositions from Sec.~\ref{QFIsection} can be used, and they can be still be taken to satisfy the quantum version of Van Trees' SPLOT approximation to the Karhunen-Lo\'{e}ve expansion.  In terms of the modal annihilation operators, the observations for the $H_0 = \mbox{range delay $\tau$ equals $\tau_{\rm min}$}$ versus $H_1 = \mbox{range delay $\tau$ equals $\tau_{\rm min}+\tau'$}$ hypothesis test are $\{\hat{a}_{R_n},\hat{a}_{I_n}\}$, where
\be
\hat{a}_{R_n} = \sqrt{\kappa}\,\hat{a}_{S_n}e^{i2\pi n\tau/T} + \sqrt{1-\kappa}\,\hat{a}_{B_n}.
\ee
Conditioned on the true hypothesis, $\{\hat{a}_{R_n},\hat{a}_{I_n}\}$ is a collection of statistically independent mode pairs that are in zero-mean, jointly-Gaussian states whose Wigner covariance matrix, given $H_k$ is true for $k = 0,1$, is 
\be
\Lambda^{(k)}_{R_nI_n} = \frac{1}{4}\!
\left[
\begin{array}{cc}
[2\kappa S^{(n)}(2\pi n/T) +2N_B + 1] {\bf I}&2\sqrt{\kappa S^{(n)}(2\pi n/T)(S^{(n)}(2\pi n/T)+1)}\,{\bf R}_{\phi^{(k)}_n}\\[.05in]
2\sqrt{\kappa S^{(n)}(2\pi n/T)(S^{(n)}(2\pi n/T)+1)}\,{\bf R}_{\phi^{(k)}_n}&[2S^{(n)}(2\pi n/T)+1]{\bf I}
\end{array} 
\right],
\label{CondxCovar}
\ee
where
\be
\phi^{(k)}_n = \left\{\begin{array}{ll}
2\pi n\tau_{\rm min}/T, & \mbox{for $k = 0$,}\\[.05in]
2\pi n(\tau_{\rm min} + \tau')/T, & \mbox{for $k =1$.}
\end{array}\right.
\ee
The desired QCB,
\be
P_e(\tau') \le e^{-\xi_{\rm QCB}(\tau')}/2,
\label{QCBgeneral}
\ee
can now be obtained using the technique from Ref.~\cite{Pirandola2008}.  When $S^{(n)}(\omega) \ll 1, \kappa \ll 1, N_B \gg 1$, we get the asymptotic result 
\be
\xi_{\rm QCB}(\tau') = \sum_n\frac{\kappa S^{(n)}(2\pi n/T)|1-e^{-i2\pi n\tau'/T}|^2}{N_B} \approx T\!\int\!\frac{{\rm d}\omega}{2\pi}\,\frac{\kappa S^{(n)}(\omega)|1-e^{-i\omega\tau'}|^2}{N_B},
\ee
where the approximation's validity is due to $T \gg 2\pi/\Delta\omega$.  Note that our derivation applies to all $S^{(n)}(\omega)$ satisfying the low-brightness condition $S^{(n)}(\omega)\ll 1$, whereas the main text assumes $S^{(n)}(\omega)/2\pi = N_Se^{-\omega^2/2\Delta\omega^2}/\sqrt{2\pi}$.  

We complete our treatment of the quantum pulse-compression radar's $\delta\tau_{\rm ZZB\text{-}QCB}$ by developing its low-SNR and high-SNR asymptotes and finding the threshold SNR where they agree.  Paralleling what was done for the classical radar, we can easily accomplish those goals.  At low SNRs we can use
\be
T\!\int\!\frac{{\rm d}\omega}{2\pi}\,\kappa S^{(n)}(\omega)|1-e^{-i\omega\tau'}|^2 \approx 2T\!\int\!\frac{{\rm d}\omega}{2\pi}\,\kappa S^{(n)}(\omega) = 2\kappa\calE,
\ee
giving us the low-SNR asymptotic behavior
\be
\delta\tau^{\rm Q}_{\rm ZZB\text{-}QCB} \approx \sqrt{\int_0^{\Delta\tau}\!{\rm d}\tau'\, \tau'\!\left(1-\frac{\tau'}{\Delta\tau}\right)\!\frac{e^{-2\,{\rm SNR}}}{2}} = \sigma_\tau e^{-{\rm SNR}}.
\label{QuantQCBlowSNR}
\ee
Likewise, at high SNRs we can use
\be
T\!\!\int\!\frac{{\rm d}\omega}{2\pi}\,S^{(n)}(\omega)|1-e^{-i\omega\tau'}|^2 \approx
T\!\int\!\frac{{\rm d}\omega}{2\pi}\,S^{(n)}(\omega)\omega^2\tau'^2 = \kappa\calE\Delta\omega^2\tau'^2,
\ee
giving us the high-SNR asymptotic behavior
\be
\delta\tau^{\rm Q}_{\rm ZZB\text{-}QCB} \approx \sqrt{\int_0^{\Delta\tau}\!{\rm d}\tau'\, \tau'\!\left(1-\frac{\tau'}{\Delta\tau}\right)\frac{e^{-{\rm SNR}\,\Delta\omega^2\tau'^2}}{2}} \approx \frac{1}{2\Delta\omega\sqrt{\rm SNR}},
\label{QuantQCBhiSNR}
\ee
which equals $\delta\tau^{\rm Q}_{\rm CRB}$, as noted in the main text.  Equating the low-SNR and high-SNR asymptotes then gives us 
\be 
{\rm SNR}^{\rm Q}_{\rm thresh} = \frac{1}{2}f\!\left(\frac{1}{2 \Delta \omega^2\sigma_\tau^2}\right).
\label{ThreshApprox}
\ee
which equals ${\rm SNR}^{\rm C}_{\rm thresh}/4$, showing---within the approximations used in getting these threshold SNRs---the quantum radar's 6\,dB lower threshold SNR than its classical radar counterpart.  The numerics used in obtaining the main text's Fig.~\ref{fig:dR} are consistent with Eq.~\eqref{ThreshApprox}, i.e., the quantum and classical radar's observed threshold SNRs are accurately predicted as is the former's  6\,dB advantage over the latter. 

To further our understanding of the quantum radar's range-accuracy advantage over classical radar, let us define
\be
\delta\tau^{\rm C}_{\rm ZZB\text{-}QCB^*} \equiv \sigma_\tau e^{-f(1/2\Delta\omega^2\sigma_\tau^2)/8},
\ee
making it our asymptotic approximation to the classical radar's rms range-delay accuracy when its SNR equals ${\rm SNR}^{\rm Q}_{\rm thresh}$, and 
\be 
\delta\tau^{\rm Q}_{\rm ZZB\text{-}QCB^*} \equiv \frac{1}{\Delta\omega\sqrt{2f(1/2\Delta\omega^2\sigma^2_\tau)}},
\ee
making it our asymptotic approximation to the quantum radar's rms range-delay accuracy when its SNR equals ${\rm SNR}^{\rm Q}_{\rm thresh}$.  These imply that the mean-squared range-delay accuracy advantage enjoyed by the quantum pulse-compression radar when operated at its threshold SNR is a factor of 
\be
(\delta\tau^{\rm C}_{\rm ZZB\text{-}QCB^*})^2/(\delta\tau^{\rm Q}_{\rm ZZB\text{-}QCB^*})^2= e^{3f(1/2\Delta\omega^2\sigma^2_\tau)/4}
 \approx \alpha \left[2\Delta \omega^2 \sigma_\tau^2\right]^{3/4},
\label{asymp_final}  
\ee
where the equality uses $f(y)=e^{f(y)}y$ and the approximation comes from  $f(y)\sim - \ln(y)$ and $\alpha$ is a fitting parameter.

Figure~\ref{fig:dR_app}(a) demonstrates the utility of Eq.~\eqref{asymp_final} for the parameters assumed in the main text's Fig.~\ref{fig:dR}, i.e., it plots
$20\log_{10}\!\left[\delta\tau^{\rm C}_{\rm ZZB}/\delta\tau^{\rm Q}_{\rm ZZB\text{-}QCB}\right]$ versus target's range uncertainty $\Delta R$ assuming that the $\Delta\omega/2\pi = 10^6\,$Hz, $|S(\omega)|^2/2\pi = e^{-\omega^2/2\Delta\omega^2}/\sqrt{2\pi\Delta\omega^2}$, and $S^{(n)}(\omega) = N_Se^{-\omega^2/2\Delta\omega^2}/\sqrt{2\pi}$.  The stars in that figure are obtained by numerical evaluation of 
$\delta\tau^{\rm C}_{\rm ZZB}$ using Eq.~\eqref{Ziv_Zakai_classical} and $P_e(\tau')$ from \eqref{cohPeGenOmega} for the classical radar, and numerical evaluation of the main text's Eq.~\eqref{Ziv_Zakai_simplified} using the Chernoff bound from \eqref{QCBgeneral} as found, for the conditional covariances given in Eq.~\eqref{CondxCovar}, from Ref.~\cite{Pirandola2008}.  These stars are in excellent agreement with the solid curve, which is obtained from Eq.~\eqref{asymp_final} with $\alpha = 0.14$.   

Figure~\ref{fig:dR_app}(b) verifies Eq.~\eqref{ThreshApprox}'s prediction that the quantum radar's threshold SNR is 6\,dB lower than that of the classical radar, i.e., it plots the classical-to-quantum threshold SNR ratio (in dB) versus the range uncertainty $\Delta R$.  The stars evaluate that threshold advantage using the numerics described for part (a) while the solid curve is the 6\,dB predicted by Eq.~\eqref{ThreshApprox}.  As in (a), we see excellent agreement between the numerical results and the analytical approximation.   
\begin{figure}[t]
    \centering
    \includegraphics[width=0.475\textwidth]{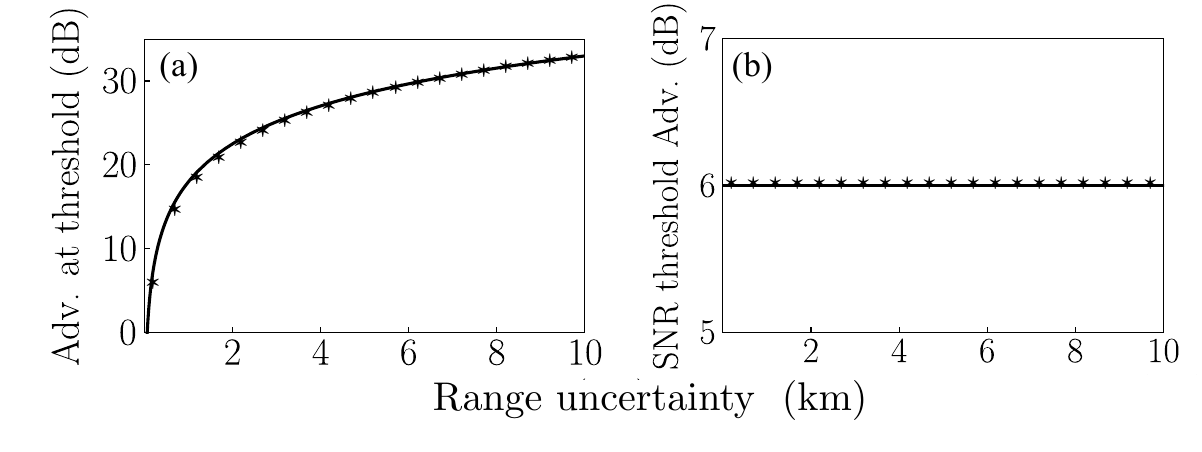}
    \caption{Performance advantage of quantum pulse-compression radar over a classical radar of the same transmitted energy and rms bandwidth.  The parameters assumed are $\Delta\omega/2\pi = 10^6\,$Hz, $|S(\omega)|^2/2\pi = e^{-\omega^2/2\Delta\omega^2}/\sqrt{2\pi\Delta\omega^2}$, and $S^{(n)}(\omega) = N_Se^{-\omega^2/2\Delta\omega^2}/\sqrt{2\pi}$, as in the main text's Fig.~\ref{fig:dR}.   
 (a) Mean-squared accuracy advantage in dB versus the target's range uncertainty in km Stars are numerical results---see text for details---while the solid curve is the approximation from Eq.~\eqref{asymp_final} with $\alpha = 0.14$.   The 6-dB advantage of the threshold SNR value. 
    \label{fig:dR_app}
    }
\end{figure}

\end{widetext}


\begin{thebibliography}{99}
\bibitem{footnote1}The field operators $\hat{E}_S(t)$ and $\hat{E}_S^\dagger(t)$, respectively, annihilate and create photons at time $t$, and their commutator is $[\hat{E}_S(t),\hat{E}_S^\dagger(t')] = \delta(t-t')$, where $\delta(\cdot)$ is the unit impulse.    Similarly, their Fourier-domain duals $\hat{A}_S(\omega)$ and $\hat{A}_S^\dagger(\omega)$, respectively, annihilate and create photons at frequency $\omega_0+\omega$, and their commutator is $[\hat{A}_S(\omega),\hat{A}_S(\omega')] = 2\pi \delta(\omega-\omega')$.
\end{thebibliography}
\begin{thebibliography}{20}
\bibitem{footnote2}Here, we have suppressed the carrier-frequency term that appeared in the main text above its Eq.~\eqref{ChirpedGaussian} because we use $\omega_0 = 0$ in the Fisher information calculation to avoid the $2\pi/\omega_0$ range ambiguities, mentioned in the main text, that would otherwise be present.
\end{thebibliography}
\end{document}